\documentclass[prb,aps,amssymb,twocolumn]{revtex4}

\usepackage{graphicx}

\begin{document}

\title{Quantum computing with antiferromagnetic spin clusters}
\author{Florian Meier$^1$, Jeremy Levy$^{2,3}$, and
Daniel Loss$^{1,3}$}
\affiliation{$^1$Department of Physics and Astronomy, University of Basel, 
Klingelbergstrasse 82, 4056 Basel, Switzerland \\
$^2$Department of Physics and Astronomy, University of Pittsburgh, 
Pittsburgh, Pennsylvania 15260 \\
$^3$Center for Oxide-Semiconductor Materials for Quantum Computation, 
Pittsburgh, Pennsylvania 15260
}
\date{\today}

\begin{abstract}
We show that a wide range of spin clusters with antiferromagnetic 
intracluster exchange interaction
allows one to define a qubit. For these spin cluster qubits, initialization,
quantum gate operation, and readout are possible using the same techniques
as for single spins. Quantum gate operation for the spin cluster qubit 
does not require control over the intracluster exchange interaction.
Electric and magnetic fields necessary to effect quantum gates need only be
controlled on the length scale of the spin cluster rather than
the scale for a single spin. 
Here, we calculate the energy gap separating the 
logical qubit states from the next excited state and the matrix
elements which determine quantum gate operation times. We discuss 
spin cluster qubits formed by one- and two-dimensional arrays of 
$s=1/2$ spins as well as clusters formed by spins 
$s>1/2$. We illustrate the advantages of spin cluster qubits for 
various suggested implementations of spin qubits and analyze 
the scaling of decoherence time with spin cluster size.
\end{abstract}


\maketitle

\section{Introduction}

During the past years, the discovery of several powerful quantum 
algorithms~\cite{bennett:00} has 
triggered substantial research efforts aiming at the implementation
of a quantum computer in a physical system. The main difficulty is that 
qubits must be prepared, manipulated, and read out with
high fidelity while decoherence is required to remain 
small.~\cite{divincenzo:95} Solid-state
implementations of qubits exploit the versatility of
nanoscale fabrication, but suffer from decoherence times which are usually 
shorter than in many quantum optics proposals.~\cite{cirac:00} 
Electron~\cite{loss:98,burkard:99,levy:01} and 
nuclear~\cite{privman:98,kane:98} 
spins have been identified as promising candidates for qubits in a 
solid state system. The main advantage of electron or nuclear spins 
is that they are natural
two state systems and that decoherence times for the spin degree of
freedom~\cite{kikkawa:97,feher:59} are usually larger than for charge 
degrees of freedom.

Here we show that a wide variety
of spin clusters are promising candidate systems for
qubits. Qubits formed by several spins have so far mainly been discussed
in the context of exchange-only quantum 
computing,~\cite{divincenzo:00,levy:02,lidar:02}
coherence-preserving qubits~\cite{bacon:01}, and quantum computing
schemes in which the requirements on the control of exchange interactions
between spins are relaxed.~\cite{benjamin:02} However, all these schemes
require control at the single-spin level, either with local 
magnetic fields~\cite{benjamin:02} or
exchange interactions.~\cite{divincenzo:00,bacon:01}
For the spin clusters considered here, control for both
magnetic fields and exchange interactions is
required only on the length scale of the spin cluster diameter. As we have
shown in Ref.~\onlinecite{meier:02}, spin chains formed by an odd number
of antiferromagnetically coupled spins $s=1/2$ allow one to define
a logical qubit. The logical state of the qubit is encoded in the
collective state of the spin cluster. 
Here we detail that this construction 
remains valid for a wide 
range of spin clusters, independent of the details of  
intracluster exchange interaction and spin placement.
Initialization and readout of the spin cluster is achieved with the
methods developed for single spins.~\cite{loss:98,burkard:99} 
The main advantage of spin clusters is that the requirements on 
spatial control can be traded for gate operation times. The scaling of
the decoherence rate with the size of the spin cluster depends on the
microscopic decoherence mechanism. While the decoherence rate induced
by fluctuating local magnetic fields increases with cluster size, we show that
magnetic dipolar interactions  for the spin cluster
qubit are smaller than for single spins. The optimum size of the spin
cluster qubit is determined by a trade-off between the increase in
gate operation times and the decoherence rate effected by local fluctuating 
magnetic fields, the decrease in magnetic dipolar interaction energy, and
the relaxed conditions on local control. 

Any quantum computation can be decomposed into a sequence of one- and 
two-qubit quantum gates.~\cite{bareno:95} For a single-spin qubit, 
the $\hat{s}_z$ eigenstates
$|\uparrow\rangle$ and $|\downarrow\rangle$ are identified as
logical basis states $|0\rangle$ and $|1\rangle$, 
respectively.~\cite{loss:98,burkard:99}
The phase shift gate can then be realized by a magnetic field
$B_z(t)$ and the one-qubit 
rotation gate $U_{\rm rot}$ by a transverse field $B_x(t)$ which rotates
$|\uparrow\rangle$ into $|\downarrow\rangle$ and vice versa.
More generally, the equations 
\begin{equation}
\langle 0| \hat{H}^\prime | 0 \rangle  = 
\langle 1| \hat{H}^\prime | 1 \rangle  \, {\rm and} \, \,
\langle 1 | \hat{H}^\prime | 0 \rangle \neq 0
\label{eq:rotcond}
\end{equation}
constitute a sufficient condition that a Hamiltonian
$\hat{H}^\prime$ induces the unitary time evolution required
for $U_{\rm rot}$. For single spins, $\hat{H}^\prime
= g \mu_B B_x(t) \hat{s}_x$ fulfills Eq.~(\ref{eq:rotcond}).
Similarly, an exchange interaction $\hat{H}_\ast = J_\ast \hat{\bf s}_1
\cdot \hat{\bf s}_2$ generates the unitary time evolution required
for the square-root of SWAP gate~\cite{loss:98} because, in the
two-qubit product basis,
\begin{equation}
\langle 10 | \hat{H}_\ast | 01 \rangle  \neq 0.
\label{eq:swapcond}
\end{equation}
In contrast to a single spin $s=1/2$, clusters formed by $n_c$
coupled spins are not intrinsically two-state systems. In order to 
prove that a logical qubit can be defined in terms of the energy 
eigenstates of a spin cluster we will (a) identify spin clusters
with a ground state doublet $\{| 0 \rangle , | 1 \rangle   \}$ 
separated from the next excited state by an energy gap $\Delta$;
(b) identify Hamiltonians $\hat{H}^\prime$ and $\hat{H}_\ast$
which satisfy Eqs.~(\ref{eq:rotcond}) and (\ref{eq:swapcond}) and, hence,
allow one to generate a universal set of quantum gates; and (c)
quantify leakage and decoherence for the spin cluster qubit.
In particular, the evaluation of the matrix elements in 
Eqs.~(\ref{eq:rotcond}) and (\ref{eq:swapcond}) and the quantification
of excitation out of the computational basis (leakage) 
requires a detailed characterization of the 
states $\{| 0 \rangle , | 1 \rangle   \}$ which is, in general,  
nontrivial.

This paper is organized as follows. In Sec.~\ref{sec:chains} we discuss the
computational basis states for spin-$1/2$ chains. 
For this simple
geometry, it is possible to derive analytical expressions for the matrix
elements in Eqs.~(\ref{eq:rotcond}) and (\ref{eq:swapcond})
for various anisotropies and 
spatially varying intracluster exchange interaction. 
Section~\ref{sec:2d} discusses the insensitivity of spin cluster qubits to 
the details of interactions within the cluster, such as the relative 
placement of spins and  the exchange strengths.
In Sec.~\ref{sec:ls}, spins with spin 
quantum numbers larger than $1/2$ are discussed. 
In Sec.~\ref{sec:conclusions}, we draw our conclusions.

\section{Spin chains}
\label{sec:chains}

For simplicity, we first consider a spin cluster qubit formed by a spin chain,
\begin{equation}
\hat{H} = \sum_{i=1}^{n_c-1} f_j [ J_\perp(\hat{s}_{j,x} \hat{s}_{j+1,x} +
\hat{s}_{j,y} \hat{s}_{j+1,y}) +J_z \hat{s}_{j,z} \hat{s}_{j+1,z}  ]
\label{eq:sch-1}
\end{equation}
where $n_c$ is odd and $J_\perp, J_z > 0$.
The real numbers $f_j>0$ account for a spatial variation of the exchange
interaction, and $J_\perp f_j$ ($J_z f_j$) denotes the transverse 
(longitudinal) exchange interaction between sites $j$ and $j+1$.

\subsection{Isotropic spin chains}
\label{sec:chains1}

For electron spins in quantum dots, the nearest neighbor exchange
is usually of the Heisenberg form,~\cite{burkard:99}  $J = J_\perp = J_z$.
We first consider $f_j \equiv 1$, 
\begin{equation}
\hat{H} = J \sum_{j=1}^{n_c-1}  \hat{\bf s}_{j} \cdot \hat{\bf s}_{j+1}
\label{eq:sch-2}
\end{equation}
with $J>0$. Note that this is an open spin chain; a closed spin chain would 
have a fourfold degenerate ground state multiplet for odd $n_c$
that would make it unsuitable for representing a single qubit.
Because the intracluster exchange interaction 
$J$ is time independent and no external
control is required, $J$ can be adjusted already during sample
growth.

\begin{figure}
\centerline{\mbox{\includegraphics[width=8cm]{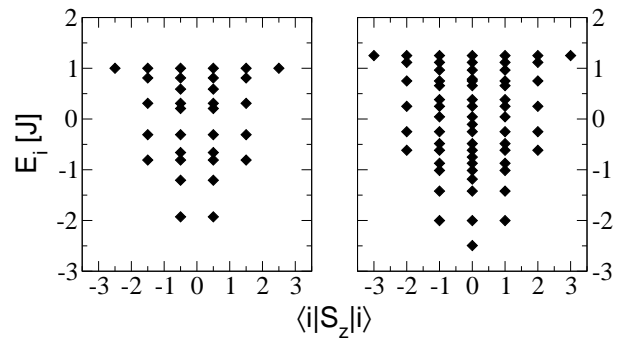}}}
\caption{Energy spectrum of an 
isotropic spin chain with $n_c=5$ (left panel) and $n_c=6$ (right panel).
Energy eigenstates are sorted according to their quantum numbers of 
$\hat{S}_z$ and their eigenenergies.
}
\label{Fig1}
\end{figure}

Spin chains have been studied in great detail during the past 
decades.~\cite{orbach:58,walker:59,lieb:61} The theoretical description
of the antiferromagnetic spin chain Eq.~(\ref{eq:sch-2}) is particularly 
challenging because the classical N{\'e}el ordered state is not an energy 
eigenstate and quantum fluctuations are pronounced. 
We define the operator of total spin,
\begin{equation}
\hat{S}_\alpha = \sum_{j=1}^{n_c} \hat{s}_{j,\alpha}
\label{eq:totspin}
\end{equation}
for $\alpha = x,y,z$.
Energy eigenstates can be 
labeled according to their quantum numbers of total spin $\hat{\bf S}$ and 
the $z$-component of total spin, 
$\hat{S}_z$, because 
\begin{equation}
[\hat{H},\hat{\bf S}^2]=[ \hat{H},\hat{S}_z]=0.
\label{eq:comrel}
\end{equation}
Due to the antiferromagnetic exchange coupling, states in 
which the total spin of the chain is minimized are energetically most 
favorable.~\cite{lieb:62}
For even $n_c$, the minimum possible spin is $S=0$, and the 
system has a nondegenerate ground state. In contrast, for odd $n_c$, there 
is a ground state doublet (Fig.~\ref{Fig1}).~\cite{lieb:62} This parity 
effect is 
well known for thermodynamic quantities.~\cite{bonner:64}
The energy gap $\Delta$ separating the
ground state doublet from the next excited state, 
\begin{equation}
 \Delta \simeq \frac{J \pi}{2} k_{\rm min} \sim   \frac{J\pi^2}{2n_c} ,
\label{eq:delta}
\end{equation}
can be estimated from the lower bound of the des Cloiseaux-Pearson spectrum
and the minimum wave vector $k_{\rm min} = \pi/n_c$ 
(Ref.~\onlinecite{nagaosa}).
Henceforth, we will restrict our 
attention to spin chains with odd $n_c$.

The requirements on a candidate system for qubits
include 
initialization of the quantum computer, a universal set of quantum
gates, decoherence times long compared to gate operation times, and
readout of the qubit.~\cite{divincenzo:95}  

\subsubsection{Definition of the spin cluster qubit}

\begin{figure}
\centerline{\mbox{\includegraphics[width=8cm]{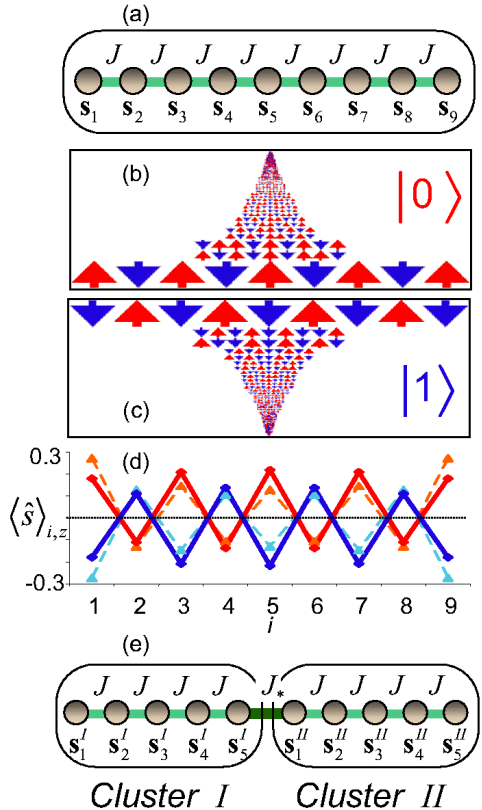}}}
\caption{Spin cluster qubit. (a) The energy eigenstates of an 
antiferromagnetic spin chain with an odd number of sites define the spin 
cluster qubit. 
(b) Wave function of $|0\rangle$ in the single-spin product basis for $n_c=9$.
The size of each configuration is proportional to the probability of
finding the corresponding product state in $|0\rangle$.
(c) Similar to (b) but for state $|1\rangle$.
(d) Spin density $\langle \psi|\hat{s}_{i,z}|\psi\rangle$ for $|\psi\rangle
= |0\rangle$ and $|\psi\rangle = |1\rangle$ (solid lines)
for constant intrachain exchange coupling. 
Dashed lines indicate the corresponding 
data for spatially varying intrachain exchange with $f_j = \sin (j\pi /n_c)$.
(e) Spin cluster qubits are coupled by a switchable interchain exchange
coupling $J_\ast(t)$. 
}
\label{Fig2}
\end{figure}

For the chain with an odd number
of sites [Fig.~\ref{Fig2}(a)], we define the spin cluster qubit in terms of 
the 
$S=1/2$ ground state doublet by 
\begin{eqnarray}
\hat{S}_z|0\rangle &=& \frac{1}{2} |0\rangle, \nonumber \\ 
\hat{S}_z|1\rangle &=& -\frac{1}{2} |1\rangle . \label{eq:qubitdef}
\end{eqnarray}
The states $\{|0\rangle,|1\rangle \}$ do not in
general have a simple representation in the single-spin product basis, but
rather are  
complicated superpositions of $n_c!/[(n_c-1)/2]![(n_c+1)/2]!$ states 
[Figs.~\ref{Fig2}(b) and (c)] as evidenced by the local magnetization density
[Fig.~\ref{Fig2}(d)]. 
The largest amplitude in this superposition corresponds
to the N{\'e}el ordered states $|\uparrow\rangle_1
|\downarrow\rangle_2 \ldots |\uparrow \rangle_{n_c}$ ($|0\rangle$) and 
$|\downarrow\rangle_1|\uparrow\rangle_2 \ldots |\downarrow \rangle_{n_c}$ 
($|1\rangle$), respectively. For $n_c=9$, the N{\'e}el configuration has 
only a $20$\% probability; the
remaining $80$\% represent quantum fluctuations [see Figs.~\ref{Fig2}(b) and 
(c)].

In spite of their complicated representation in the single-spin product basis, 
$|0\rangle$ and $|1\rangle$ are in many respects very similar 
to the states $|\uparrow\rangle$ and $|\downarrow\rangle$ of a single spin
and, hence, define the computational basis 
for universal quantum computing with \emph{spin cluster qubits}. 
The reason for this is that $\{|0\rangle,|1\rangle \}$ belong to one 
$S=1/2$ doublet  
such that 
\begin{equation}
\hat{S}^-|0\rangle = |1\rangle, \,\,\,\, 
\hat{S}^+|1\rangle = |0\rangle, \label{eq:ch-doublet}
\end{equation}
where $\hat{S}^{\pm}
= \hat{S}_x \pm i \hat{S}_y$ are the spin ladder operators of the 
spin chain. 

\subsubsection{Initialization}

Initialization of the spin cluster qubit can be achieved by cooling
in a magnetic field with $g \mu_B B_z \ll \Delta$ to temperatures
$T \ll g \mu_B B_z/k_B$. The spin cluster will relax
to $|0\rangle$ within the spin relaxation time. Because thermalization
is typically slow, the preparation of a given initial state could
be facilitated by the measurement of the spins within the clusters, possibly
followed by local operations. In this way, a state which is energetically
close to the ground state could be prepared.

\subsubsection{Quantum Gate operation}

The one-qubit phase shift gate $U_\phi$, one-qubit rotation
gate $U_{\rm rot}$, and CNOT gate $U_{\rm CNOT}$ constitute a universal set of 
quantum gates. According to Eq.~(\ref{eq:ch-doublet}), a
magnetic field that is constant over the spin cluster qubit acts on the
states of the spin cluster qubit in the same way as on a single-spin 
qubit.~\cite{meier:02} 
Constant magnetic fields $B_z$ ($B_x$) effect the one-qubit
phase shift (one-qubit rotation) quantum gate 
without leakage. Because 
$|0\rangle$ and $|1\rangle$ are degenerate and
separated from the next excited state by $\Delta$, one-qubit 
quantum gates can be realized with high fidelity also by any spatially 
varying magnetic field for which 
$|\langle 1|\sum_{j=1}^{n_c} g_j \mu_B B_{j,x} \hat{s}_{j,x}|0\rangle| \neq 0$
($U_{\rm rot}$)  and
$\langle 0|\sum_{j=1}^{n_c} g_j \mu_B B_{j,z} \hat{s}_{j,z}|0\rangle \neq 0$
($U_\phi$), 
respectively. Here, $g_j$ is the electron spin $g$ factor at site $j$.
Such spatially varying fields can potentially cause leakage.
However, if $|g_j \mu_B B_j|/\Delta \ll 1$ and if all $B_j$ are switched on 
and off adiabatically, i.e., 
on time scales long compared to $h/\Delta$, the quantum gate fidelity
is close to $100$\% as we will discuss in detail in Sec.~\ref{sec:chains2}.

For the CNOT gate, one requires an exchange interaction $\hat{H}_{\ast}$ 
between one or several spins of 
neighboring spin cluster qubits I and II which can be switched on 
and off, e.g., by electrical gates.~\cite{loss:98} 
The simplest case of an exchange interaction
between the outermost spins of neighboring clusters [Fig.~\ref{Fig2}(e)],
$\hat{H}_{\ast}=J_{\ast}(t) \hat{\bf s}_{n_c}^{\rm I} \cdot 
\hat{\bf s}_{1}^{\rm II}$ translates into an isotropic
exchange interaction also for the two-qubit product basis. This remains
true for any  $\hat{H}_{\ast}$ of the form
\begin{equation}
\hat{H}_{\ast}=J_{\ast}(t) \hat{\bf s}_{n_c}^{\rm I} \cdot 
\hat{\bf s}_{1}^{\rm II}
+ J_{\ast}(t) \sum_{j=1}^{n_c-1} (v_j^{\rm I}  \hat{\bf s}_{j}^{\rm I} \cdot
 \hat{\bf s}_{j+1}^{\rm I}+ v_j^{\rm II}  \hat{\bf s}_{j}^{\rm II} \cdot
 \hat{\bf s}_{j+1}^{\rm II}).
\label{eq:coup-ham}
\end{equation}
Here, the factors $v_j$ allow for a spatial 
variation of the intracluster exchange coupling constants 
during gate operation, where
$|v_j|<1$ and $| J_\ast| \ll J$ is the limit relevant for experiments. 
$\hat{H}_\ast$ will in general not only couple states
within the two-qubit product basis $\{|00\rangle, |01\rangle,|10\rangle,
|11\rangle \}$, but will also lead to leakage. 
As long as $J_{\ast}(t)$ changes adiabatically  
and $|J_{\ast} (t)| \ll \Delta$ for all
times, leakage remains small. It should be noted that this adiabaticity 
condition also holds for single electrons in quantum dots where, however,
the energy gap $\Delta$ is usually larger than for the spin cluster qubit.
The action of
$\hat{H}_{\ast}$ can then be described by an
effective Hamiltonian in the two-qubit product basis, 
\begin{eqnarray}
\hat{H}_{\ast} &=& J_{\ast z}(t)  \hat{S}^{\rm I}_z   \hat{S}^{\rm II}_z 
+ \frac{J_{\ast \perp}(t)}{2} ( \hat{S}^{{\rm I}+}   \hat{S}^{{\rm II}-}
+ \hat{S}^{{\rm I}-}   \hat{S}^{{\rm II}+})  \nonumber \\
&& \hspace*{3cm} + J_{o}(t) {\bf 1}
 , \label{eq:cnot1}
\end{eqnarray}
where the roman numbers label the spin clusters, and 
\begin{eqnarray}
J_{\ast z}(t) &=&
4 J_{\ast}(t) |_{\rm I}\langle 0|\hat{s}_{n_c,z}^{\rm I}|0\rangle_{\rm I}|
|_{\rm II}\langle 0|\hat{s}_{1,z}^{\rm II}|0\rangle_{\rm II}|,
  \nonumber \\
J_{\ast \perp}(t) &=& 4  J_{\ast}(t) 
|_{\rm I}\langle 1|\hat{s}_{n_c,x}^{\rm I}|0\rangle_{\rm I}| 
|_{\rm II}\langle 1|\hat{s}_{1,x}^{\rm II}|0\rangle_{\rm II}|, 
\label{eq:ef-coupl}\\
J_o(t) &=& J_{\ast}(t) [_{\rm I}\langle 0|\sum_{j=1}^{n_c-1} v_j^{\rm I}  
\hat{\bf s}_{j}^{\rm I} \cdot
 \hat{\bf s}_{j+1}^{\rm I}|0\rangle_{\rm I}  \nonumber \\ && \hspace*{1cm} +
_{\rm II}\langle 0|\sum_{j=1}^{n_c-1} v_j^{\rm II} \hat{\bf s}_{j}^{\rm II} 
\cdot \hat{\bf s}_{j+1}^{\rm II}|0\rangle_{\rm II}]. \nonumber 
\end{eqnarray} 
For the derivation of Eq.~(\ref{eq:cnot1}), see Appendix~\ref{sec:ap1}.

Because, for the isotropic chain,  
\begin{equation}
_{\rm I}\langle 0|\hat{s}_{n_c,z}^{\rm I}
|0\rangle_{\rm I} =|_{\rm I}\langle 1|\hat{s}_{n_c,x}^{\rm I}
|0\rangle_{\rm I}|,
\label{eq:cnot1b}
\end{equation} 
the coupling $\hat{H}_{\ast}$ is  isotropic also
in the two-qubit product basis and acts on the states 
$|0\rangle$
and $|1\rangle$ of neighboring spin chains {\it in the same way as an
isotropic exchange interaction between two single spins}. 

An explicit switching sequence for the CNOT gate based on single-qubit
rotations and the unitary time evolution governed by the exchange
interaction in Eq.~(\ref{eq:cnot1}) is~\cite{meier:02}
\begin{eqnarray}
 && U_{\rm CNOT}\sim  e^{-i \pi S_{y}^{\rm II}/2}
e^{i 2 \pi {\bf n}_1 \cdot {\bf S}^{\rm I}/3} e^{i 2 \pi {\bf n}_2 
\cdot {\bf S}^{\rm II}/3}
 U_{\ast}(\pi/2)  \nonumber \\ && \hspace*{0.5cm}
\times  e^{i \pi S_{y}^{\rm I}} U_{\ast} (\pi/2)
e^{-i \pi S_{x}^{\rm I}/2} e^{-i \pi S_{x}^{\rm II}/2}  
e^{i \pi S_{y}^{\rm II}/2} \label{eq:cnot2}
\end{eqnarray} 
for  the general case where  $J_{\ast z} \neq  J_{\ast \perp}$. 
Here, ${\bf n}_1= (1,-1,1)/\sqrt{3}$ and ${\bf n}_2= (1,1,-1)/\sqrt{3}$, and
we have defined the unitary time evolution operator 
$U_{\ast} (\pi/2) = \hat{\rm T}_t 
\exp  \left(-i \int dt \, \hat{H}_{\ast} /\hbar\right)$,
with $-\int dt \, J_{\ast \perp} (t) /\hbar = \pi/2$.

The gate operation time for $U_\ast(\pi/2)$ is limited from below by
$h/16 J_\ast 
|_{\rm I}\langle 1|\hat{s}_{n_c,x}^{\rm I}|0\rangle_{\rm I}| 
|_{\rm II}\langle 1|\hat{s}_{1,x}^{\rm II}|0\rangle_{\rm II}|$, where
$J_{\ast}$ is the maximum value of the exchange coupling. Matrix 
elements such as $|_{\rm I}\langle 1|\hat{s}_{n_c,x}^{\rm I}|
0\rangle_{\rm I}|$ decrease with increasing $n_c$, which leads to an
increase in gate operation time. For realistic 
parameters and small $n_c$ (see Sec.~\ref{sec:chains6} below), $J_\ast$
is limited by experimental constraints rather than the condition
$J_\ast \ll \Delta$. Then, the increase in gate operation time compared
to single-spin qubits is $1/(4 |_{\rm I}\langle 1|\hat{s}_{n_c,x}^{\rm I}
|0\rangle_{\rm I}| |_{\rm II}\langle 1|\hat{s}_{1,x}^{\rm II}|
0\rangle_{\rm II}|)$ and depends only on the matrix elements of
the spin operators. Similarly, for a given magnetic field, the
increase of the time required for a one-qubit rotation depends only 
on the matrix elements $|\langle 1|\hat{s}_{j,x}|0\rangle|$.

\subsubsection{Decoherence}

For spin clusters, decoherence usually
is faster than for single spins. The scaling
of the decoherence time $\tau_\phi$ with system size depends on the 
microscopic decoherence mechanism. For electron spins in quantum dots,
fluctuating fields and nuclear spins have been
identified as dominant 
sources.~\cite{loss:98,burkard:99,khaetskii:02,merkulov:02,schliemann:02,khaetskii:03,sousa:03}
In the following, we discuss two scenarios for decoherence in which
(i) a fluctuating field couples to the total magnetic moment of the
spin cluster qubit, and (ii) independent fluctuating fields act on each
spin of the spin cluster qubit individually.

In order to obtain analytical estimates
for the scaling of $\tau_\phi$ with $n_c$, we restrict our analysis to
a phenomenological model in which decoherence is effected by a fluctuating
classical field $\hbar b(t)$. In case (i), 
\begin{equation}
\hat{H}^B = \hbar b(t) \hat{S}_z.
\label{eq:dec1}
\end{equation}
The decoherence rate is obtained from~\cite{blum:96}
\begin{eqnarray}
\frac{1}{\tau_\phi} &=& \pi \sum_{|k\rangle \neq |0\rangle}
|\langle k|\hat{S}_z|0\rangle|^2 C(E_k-E_0) \nonumber \\
&& 
+ \pi \sum_{|k\rangle \neq |1\rangle}
|\langle k|\hat{S}_z|1\rangle|^2 C(E_k-E_1) \label{eq:dec3} \\
&&  + \pi (\langle 0|\hat{S}_z|0\rangle
-  \langle 1|\hat{S}_z|1 \rangle)^2 C(0), \nonumber
\end{eqnarray} 
where
\begin{equation}
C(E) = \frac{1}{2\pi} \int_{-\infty}^{\infty} dt \, e^{iEt/\hbar}
\langle b(t) b(0) \rangle 
\end{equation}
is the spectral density of the random field in Eq.~(\ref{eq:dec1}). 
Because of Eq.~(\ref{eq:qubitdef}), 
we find
that the decoherence rate $1/\tau_\phi=\pi C(0)$ is  independent of 
$n_c$. This result, which is in stark contrast to the standard result
that the decoherence rate increases with system size, 
can be traced back to the fact that in Eq.~(\ref{eq:dec1}) only the total 
magnetic moment couples
to the fluctuating field and the magnetic moment $\pm g \mu_B/2$ of 
the spin cluster qubit is identical to the one of a single spin. 
Similarly, for a non-diagonal coupling $\hat{H}^B = \hbar b(t) \hat{S}_x$, 
from Eq.~(\ref{eq:dec3}) we find $1/\tau_\phi=\pi C(0)/2$.

Decoherence due to the coupling to nuclear spins is a complicated theoretical 
problem in its own right.~\cite{khaetskii:02,merkulov:02,schliemann:02,khaetskii:03,sousa:03}
In order to obtain a heuristic
estimate for the scaling of the decoherence time with $n_c$ for this
decoherence scenario, 
we consider fluctuating classical fields which act independently on 
each site of the spin cluster,
\begin{equation}
\hat{H}^B = \sum_{j=1}^{n_c} \hbar b_j(t) \hat{s}_{j,z},
\label{eq:dec2b}
\end{equation}
where $\langle b_i(t) b_j(0)  \rangle \propto \delta_{ij}$. 
For Gaussian white noise with
\begin{equation}
\langle b_i(t) b_j(0)\rangle
= 2 \pi \gamma^B \delta_{ij} \delta (t),
\label{eq:dec2}
\end{equation} 
$1/\tau_\phi \simeq  n_c \pi\gamma^B$ scales linearly with $n_c$, i.e.,
if fluctuating fields act independently on the individual spins of the
cluster, the decoherence rate increases with $n_c$. This result
for the scaling of $1/\tau_\phi$ with system size agrees with the standard 
result for $n_c$ qubits which are subject to independently fluctuating 
fields.~\cite{unruh:95} Note that an increase in the decoherence
rate by a factor $n_c$ implies that the probability for the state of
the spin cluster qubit to remain unaffected by the fluctuating fields 
for a certain time $t$ decreases exponentially with $n_c$ 
(Ref.~\onlinecite{unruh:95}).

Magnetic dipolar interactions are another source of decoherence. 
Consider two single-spin qubits $\hat{\bf s}_1$
and $\hat{\bf s}_2$ with coordinate vectors ${\bf r}_1$ and
${\bf r}_2$, respectively, that are coupled by the
magnetic dipolar interaction,
\begin{equation}
\hat{H}_{\rm dip} = \frac{\mu_0 (g \mu_B)^2}{4\pi} \frac{\hat{\bf s}_1
\cdot \hat{\bf s}_2 - 3(\hat{\bf s}_1 \cdot {\bf e}_r)
(\hat{\bf s}_2 \cdot {\bf e}_r)}{d^3},
\label{eq:dip1}
\end{equation}
where ${\bf r}={\bf r}_1-{\bf r}_2$, $d=|{\bf r}|$ is the distance between
the single-spin qubits, and ${\bf e}_r = {\bf r}/|{\bf r}|$.
In first order  
perturbation theory, the dipolar interaction leads to an energy
shift $\pm E_{\rm dip}$  for the product states with parallel and antiparallel 
spin configurations, respectively. For lateral quantum dots with spin
quantization axis perpendicular to ${\bf e}_r$,
\begin{eqnarray}
E_{\rm dip} &=& \langle \uparrow, \uparrow| 
\hat{H}_{\rm dip} |\uparrow,\uparrow\rangle \nonumber \\
&=&  \frac{\mu_0(g \mu_B)^2}{4\pi}  \frac{\langle \uparrow,\uparrow| 
\hat{\bf s}_1
\cdot \hat{\bf s}_2|\uparrow,\uparrow\rangle}{d^3} = 
\frac{\mu_0(g \mu_B)^2}{16\pi d^3}, 
\label{eq:dip2}
\end{eqnarray} 
because the expectation value of $(\hat{\bf s}_1 \cdot {\bf e}_r)
(\hat{\bf s}_2 \cdot {\bf e}_r)$ vanishes identically.
Similarly, we find
\begin{eqnarray}
\langle \downarrow,\downarrow| 
\hat{H}_{\rm dip} |\downarrow,\downarrow \rangle &=& +E_{\rm dip}, \nonumber \\
\langle \downarrow,\uparrow| \hat{H}_{\rm dip} |
\downarrow,\uparrow \rangle &=& -E_{\rm dip},\label{eq:dip3} \\
\langle \uparrow,\downarrow| \hat{H}_{\rm dip} |\uparrow,\downarrow \rangle 
&=& -E_{\rm dip}. \nonumber
\end{eqnarray}
The deterministic phase shift due to the different dipolar energies
for parallel and antiparallel spin configurations could be accounted for at
the end of a quantum algorithm
if no other error sources were present. However, it leads to correlated errors 
when spins are coupled to an environment that 
induces, e.g., spin flip errors. While $E_{\rm dip}$ is too small to induce
errors of order $10^{-4}$ for electron spins in quantum dots, dipolar
interactions are large for, e.g., P dopants in a Si matrix 
(see Sec.~\ref{sec:chains6}). 
We show
next that dipolar interactions for spin cluster qubits are smaller than
for single-spin qubits. For definiteness, we consider two clusters with
spins at sites ${\bf r}_j^{\rm I} = jd{\bf e}_y$ and ${\bf r}_j^{\rm II} = 
(j+n_c)d{\bf e}_y$ for clusters I and II, respectively, with
$j = 1,2,\ldots,n_c$ [Fig.~\ref{Fig2}(c)]. The intracluster dipolar 
interactions lead to an unimportant energy shift
that is identical for both $|0\rangle$ and $|1\rangle$,
\begin{eqnarray}
&& \sum_{i > j} \frac{\mu_0 (g \mu_B)^2}{4\pi d^3} \langle 0| 
\frac{\hat{\bf s}_i \cdot \hat{\bf s}_j - 3 \hat{s}_{i,y} 
\hat{s}_{j,y}}{(i-j)^3} |0\rangle \nonumber \\
&& \hspace*{1cm} = \sum_{i > j} \frac{\mu_0 (g \mu_B)^2}{4\pi d^3} \langle 1| 
\frac{\hat{\bf s}_i \cdot \hat{\bf s}_j - 3 \hat{s}_{i,y} 
\hat{s}_{j,y}}{(i-j)^3} |1\rangle
\label{eq:dip4}
\end{eqnarray} 
because of symmetry.
The interqubit dipolar interaction,
\begin{equation}
\hat{H}_{\rm dip}^{\rm sc} = \frac{\mu_0 (g \mu_B)^2}{4\pi d^3}
\sum_{i, j} \frac{\hat{\bf s}_i^{\rm I} \cdot \hat{\bf s}_j^{\rm II} - 
3 \hat{s}_{i,y}^{\rm I} 
\hat{s}_{j,y}^{\rm II}}{(i-j+n_c)^3},
\label{eq:dip5}
\end{equation}
gives rise to an energy shift of the states $|00\rangle$ and 
$|11\rangle$ relative to $|01\rangle$ and $|10\rangle$, where $|0\rangle$ 
and $|1\rangle$ denote the logical basis of the spin cluster qubit,
\begin{eqnarray}
\langle 00| \hat{H}_{\rm dip}^{\rm sc} |00\rangle = \langle 11| 
\hat{H}_{\rm dip}^{\rm sc} |11 \rangle &=& E_{\rm dip}^{\rm sc}, \nonumber \\
\langle 01| \hat{H}_{\rm dip}^{\rm sc} |01\rangle = 
\langle 10| \hat{H}_{\rm dip}^{\rm sc} |10 \rangle &=& - E_{\rm dip}^{\rm sc},
\label{eq:dip6}
\end{eqnarray}
similarly to single-spin qubits. Evaluating the matrix elements numerically,
we find that the characteristic dipolar energy $E_{\rm dip}^{\rm sc}$
decreases with increasing $n_c$ We find 
$E_{\rm dip}^{\rm sc}/E_{\rm dip} = 0.42, 0.25, 0.16$, and $0.12$
for $n_c = 3,5,7$, and $9$, respectively, where $E_{\rm dip}$ is the
dipolar energy for single-spin qubits. This shows that decoherence effected 
by 
magnetic dipolar interactions is indeed smaller for spin cluster qubits
than for single spins.

\subsubsection{Readout}

Readout of the spin cluster qubit is achieved by measuring
the individual spins within the cluster ($\hat{s}_{i,z}$) or
the state of the total spin of the cluster ($\hat{S}_z$). The measurement
of individual spins still is a considerable experimental challenge.
However, as has been shown theoretically,~\cite{loss:98,recher:00} measurement
of single spins is feasible via charge degrees of freedom. More specifically,
the state of a single spin on a quantum dot can be detected by
a current flowing between spin polarized leads that are tunnel coupled
to the quantum dot.~\cite{recher:00}

If an experimental technique is established that allows one to measure
the state of a single spin, it will also be possible to measure
the state of a spin cluster qubit by measurement of all spins of
the cluster.  Because $\langle 0|\sum_{i=1}^{n_c}
\hat{s}_{i,z} |0 \rangle = 1/2$ and  $\langle 1|\sum_{i=1}^{n_c}
\hat{s}_{i,z} |1 \rangle = -1/2$, this 
will allow one to unambiguously determine the state of the cluster
qubit. However, the state of the cluster determines the local spin
density at each site [Fig.~\ref{Fig2}(d)], and a probabilistic readout is
possible also by measurement of
single spins only. For example, for $n_c=9$, if the measurement of the
central spin of the chain yields $+1/2$, the spin cluster
qubit has been in state $|0\rangle$ with a probability of $70$\%.
A selective readout of several spins of the spin cluster qubit would
also reduce the requirements on the readout sensitivity. For example,
the sublattice spin $\langle 0| \hat{s}_{1,z} +  \hat{s}_{3,z}
+ \hat{s}_{5,z} +  \hat{s}_{7,z} +  \hat{s}_{9,z} |0\rangle \simeq 1$
for $n_c=9$ 
is twice as large as the total magnetic moment $\langle 0 | \hat{S}_z |0
\rangle$ and could be measured after separating the electron spins
at even and odd sites of the cluster. This illustrates that readout
of single spins is sufficient but not necessary to measure the
state of the spin cluster qubit.

Spin readout can also be effected by optical means. A scheme for
gate operation based on Pauli blocking has been suggested
recently.~\cite{pazy:03} In a similar way, the creation of a
spin polarized exciton by a circularly polarized laser beam and the 
subsequent obervation of the reemitted photon would allow one
to determine the spin state of a single electron. The underlying
principle is that, for a quantum dot
with an electron spin pointing up, Pauli's exclusion principle prohibits
the creation of an exciton  in which the electron with spin up
occupies the same orbital state. The creation of an exciton, which
can be observed by the photon emitted on recombination, is possible only
if the electron spin in the quantum dot points opposite to the one of
the exciton created by the laser beam. However, for qubits based on
spins in a single quantum dot, the minimum laser focus on the order of
the wavelength puts severe constraints on an optical readout scheme.
These constraints are relaxed for spin cluster qubits where it is
sufficient to have a laser spot size with a diamater smaller than the
size of the spin cluster qubit.

\subsection{Varying exchange constants}
\label{sec:chains5}

The formation of a spin cluster qubit from 
an odd number of antiferromagnetically 
coupled spins requires little
control over intracluster exchange constants.
Although both the energy gap $\Delta$ and matrix elements such as 
$\langle 1|\hat{s}_{j,x}|0\rangle$ 
depend on the spatial variation of exchange constants, the general 
principle of assembling several spins into a cluster qubit remains valid.

In order to demonstrate the robustness of our spin cluster qubit against
a variation of exchange constants, we return to the isotropic spin
chain but now allow for varying $f_j$ in Eq.~(\ref{eq:sch-1}).  Because
the isotropic spin chain still exhibits a $S=1/2$ ground state 
doublet,~\cite{lieb:62} quantum computing is possible as discussed 
for $f_j \equiv 1$ in Sec.~\ref{sec:chains1}. From an experimental point 
of view, a priori knowledge of the factors $f_j$ is not necessary for quantum
computing. Rather, the relevant matrix elements such as 
$|\langle 1|\hat{s}_{j,x}|0\rangle|$ can be determined experimentally. 
Similarly, a quantum computer could even be assembled from spin cluster 
qubits which are not identical. 

If the exchange constants can be
controlled during sample growth, the properties of the spin cluster
qubit can be engineered to a certain extent. For 
clusters with centrosymmetric exchange constants, the time required to 
perform the square-root of swap gate $U_\ast(\pi/2)$ for given $J_\ast$ 
increases with $1/|\langle 0|\hat{s}_{1,z}|0\rangle|^2$. For $n_c=9$ and
$f_j \equiv 1$,
$|\langle 0|\hat{s}_{1,z}|0\rangle| \simeq 0.18$ corresponding
to an increase in the gate operation time for $U_\ast(\pi/2)$
by a factor
$(0.5/0.18)^2 \simeq 7.7$ compared to the single-spin qubit for given
$J_\ast$. However, by tuning the outermost couplings to small values
$f_1 =f_{n_c-1} \ll \min_{j=2,\ldots,n_c-2}f_j$, the spin density at
the outermost sites increases and approaches $1/3$ [see dashed lines
in Fig.~\ref{Fig2}(d) and Appendix~\ref{sec:ap2} for a proof of this
statement]. Although the energy gap $\Delta$ also decreases and is
approximated by $J f_1$ in this limit, a trade-off between the increasing
matrix elements and the decreasing energy gap would allow one to decrease
gate operation times compared to the chain with spatially constant exchange
coupling.

\subsection{Leakage}
\label{sec:chains2}

We next discuss in more detail the one qubit
rotation gate induced by a transverse magnetic field in order to quantify
leakage. A related analysis for the CNOT gate has been reported in 
Ref.~\onlinecite{meier:02}.

Because $\hat{S}_x|0\rangle = (\hbar/2) 
|1\rangle$ and $\hat{S}_x|1\rangle = (\hbar/2)  |0\rangle$, a magnetic
field that is uniform over the spin cluster acts on the spin cluster qubit
in the same way as on a single spin $s=1/2$. In particular, the 
Hamiltonian
\begin{equation}
\hat{H}^\prime = g \mu_B B_x (t) \hat{S}_x 
\label{eq:rot-p1}
\end{equation}
induces a coherent rotation from $|0\rangle$ to $|1\rangle$ without leakage,
as implied by quantum mechanical selection rules.
The gate operation time for a rotation by $\phi$
is determined by $\int_0^t dt^\prime g \mu_B 
B_x (t^\prime)/\hbar = \phi$ and is identical to the one for single spins.
In contrast, the one-qubit rotation effected by a spatially varying 
magnetic field
\begin{equation}
\hat{H}^\prime = \sum_{j=1}^{n_c} g_j \mu_B B_{j,x}(t) \hat{s}_{j,x}
\label{eq:rot-p2}
\end{equation}
will in general lead to leakage because of finite matrix
elements $\langle i| \hat{H}^\prime |0 \rangle \neq 0$  and
$\langle i| \hat{H}^\prime |1 \rangle \neq 0$ coupling the
computational basis to higher excited states 
$|i\rangle \neq |0\rangle, |1\rangle$. 
The adiabatic theorem guarantees that  
leakage remains small if the Fourier transform of
$B_{j,x}$ vanishes for frequencies larger than $\Delta/\hbar$. Even
if this adiabaticity requirement is not met, admixing of higher excited 
states to $\{|0\rangle, |1\rangle \}$ is controlled by the parameters
$g_j \mu_B B_{j,x} /\Delta$ and remains small if 
$|g_j \mu_B B_{j,x}| \ll \Delta$ for all $j$. 

\begin{figure}
\centerline{\mbox{\includegraphics[width=8cm]{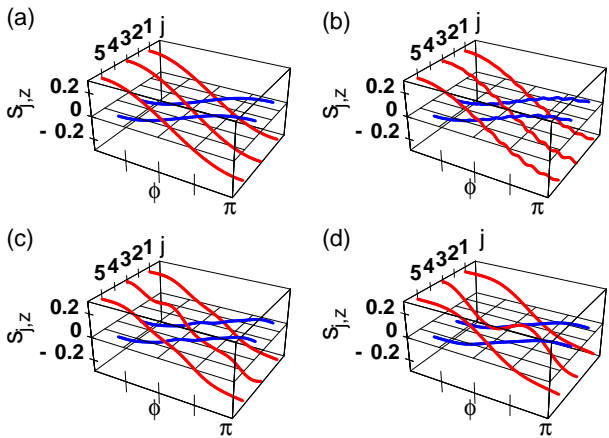}}}
\caption{Local spin density for all sites $j$ of a spin chain with 
$n_c=5$ 
as a function of time ($\phi \propto t$) during the one-qubit rotation
gate. (a) A magnetic field constant over the cluster, $\hat{H}^\prime
= g \mu_B B_x \hat{S}_x$, coherently rotates $|0\rangle$ into $|1\rangle$
without leakage. Here, $\phi = g \mu_B B_x t/\hbar$. The coherent 
rotation is evidenced by the in-phase rotation of all spins.
(b)--(d) An inhomogeneous magnetic field, $\hat{H}^\prime
= g \mu_B B_{3,x} \hat{s}_{3,x}$, effects the one-qubit rotation gate
with high gate fidelity if $|g \mu_B B_{3,x}| \ll \Delta$. 
Here, $\phi = g \mu_B B_{3,x} t (2 |\langle 1| \hat{s}_{3,x}|0\rangle|)/\hbar$.
The gate fildelity decreases from $99.8$\% to $93.4$\% and $78.5$\%
for increasing $g \mu_B B_{3,x}=0.1 J$ (b), $0.5 J$ (c), and $1 J$ (d),
respectively. In the local spin density, leakage is evidenced by 
high-frequency oscillations of neighboring spins, i.e., the excitation of 
magnons.
}
\label{Fig3}
\end{figure}

\begin{figure}
\centerline{\mbox{\includegraphics[width=8cm]{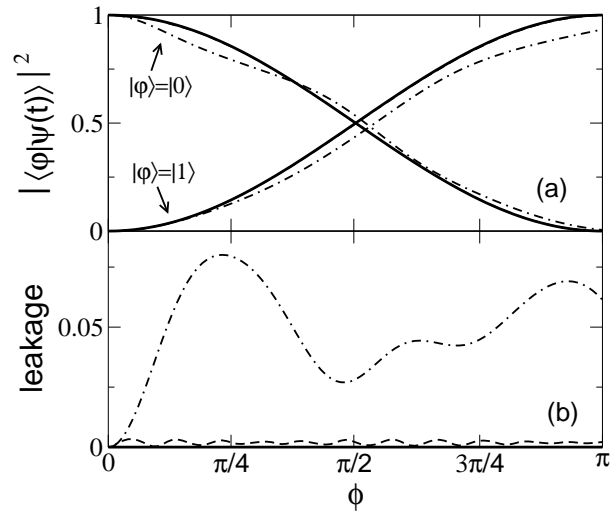}}}
\caption{Time evolution of a state $|\psi (t) \rangle$ with
$|\psi(0)\rangle = |0\rangle$ 
during the one-qubit rotation induced by a spatially 
constant magnetic field (solid line) and a magnetic field acting only on
the central spin of the cluster (dashed and dashed-dotted lines) for
$n_c=5$ as a function of time, $\phi \propto t$.
The constant magnetic field effects the quantum gate with $100$\% fildelity.
In order to illustrate the effect of leakage with increasing Zeeman energy 
for a spatially inhomogeneous field, $\hat{H}^\prime
= g \mu_B B_{3,x} \hat{s}_{3,x}$, we assume that $\hat{H}^\prime$ is switched
on instantaneously at $t=0$. (a) For $g \mu_B B_{3,x} = 0.1 J$ (dashed line),
the inhomogeneous magnetic field still effects the one-qubit rotation
with $99.8$\% fidelity. (b) Leakage out of the computational basis,
$1-(|\langle 0|\psi(t) \rangle|^2 + |\langle 1|\psi(t) \rangle|^2)$, 
remains smaller than $0.3$\%. In contrast, for $g \mu_B B_{3,x} = 0.5 J$ 
(dashed-dotted lines),
i.e., comparable to $\Delta \simeq 0.72 J$, the fidelity is only $93.4$\%
and leakage is of order $7$\%. 
}
\label{Fig4}
\end{figure}

In the following, we concentrate on $n_c=5$ spins. As shown
in Fig.~\ref{Fig3}(a) 
and Fig.~\ref{Fig4}, a magnetic field constant over
the cluster coherently rotates $|0\rangle$ into
$|1\rangle$. This is also evidenced by the in-phase rotation of all spins. 
In order
to illustrate that also a spatially inhomogeneous field can induce the
one-qubit rotation with high gate fidelity, we now consider a magnetic field 
$B_{3,x}$ acting only
on the central spin ($j=3$) of the cluster. The field is switched on 
instantaneously at $t=0$. For $t>0$, the time evolution is then governed
by the sum of the time-independent Hamiltonian of the spin cluster,
$\hat{H}$, and the 
Zeeman Hamiltonian $\hat{H}^\prime = g \mu_B B_{3,x} \hat{s}_{3,x}$. 
In Fig.~\ref{Fig4}(a), for an initial state $|\psi(t=0)\rangle = |0\rangle$, 
we plot the 
projection of the state $|\psi(t)\rangle$ onto the qubit basis states
for $g \mu_B B_{3,x}/J= 0.1$ (dashed line, coinciding with the solid line on 
this scale) and $0.5$ (dashed-dotted line), 
respectively. The time evolution is obtained by numerical integration
of the Schr{\"o}dinger equation.
For small $g \mu_B B_{3,x} \ll \Delta \simeq 0.72 J$, the 
spatially inhomogeneous field rotates $|0\rangle$ into $|1\rangle$ with
high fidelity. The gate fidelity $|\langle 1 |
U_{\rm rot}| 0 \rangle|^2$ decreases from $99.8$\% ($g \mu_B B_x/J= 0.1$)
to $93.4$\% for $g \mu_B B_x = 0.5 J$, where the typical energy scale of
$\hat{H}^\prime$ becomes comparable to $\Delta$. Here, $U_{\rm rot} = 
\hat{\rm T}_t \exp \left( -i \int_0^{t_{max}} dt 
(\hat{H} + \hat{H}^\prime)/\hbar \right)$ 
with $t_{max} = h/(4|\langle 1| \hat{s}_{x,3} |0\rangle| g \mu_B B_x )$ 
describes the time evolution during a $\pi$-rotation. 

The 
decrease in gate fidelity with increasing $B_{3,x}$ can be understood from
the local spin densities [Figs.~\ref{Fig3}(b), (c), and (d)]. 
Although only the central spin is acted on by $B_{3,x}$,
for $|g \mu_B B_{3,x}| \ll \Delta$  all spins of the spin cluster corotate 
with the
central spin due to the exchange field. The condition $|g \mu_B B_{3,x}| \ll 
\Delta$ guarantees that the externally induced rotation of the central
spin is sufficiently slow that all spins of the chain corotate in phase.
For $|g \mu_B B_{3,x}|$ of order  $\Delta$,
the rotation of the central spin induced by $B_{3,x}$
is too fast for the remaining spins of the chain to follow in phase [see, 
e.g., 
Fig.~\ref{Fig3}(c) for short times, $\phi \ll 1$]. The spins of the chain no
longer rotate in phase and magnons are excited [Fig.~\ref{Fig4}(b)].
Quantum gate
operation probes the spin dynamics in real time and, hence, may provide new 
insight into the low energy physics of spin chains. 
That~leakage~is~controlled~by~the~parameter $|g \mu_B B_{3,x}|/\Delta$
can be traced back to the existence of a ground state doublet. 

To illustrate this point, we next contrast our results for a 
system with antiferromagnetic exchange coupling and a ground state doublet
with a ferromagnetic chain with exchange
coupling $J<0$ in Eq.~(\ref{eq:sch-2}). For the ferromagnetic chain, 
the ground state has degeneracy
$n_c + 1$.  A computational basis could be defined also in terms
of a subset of the ground state multiplet for $J<0$, e.g., by the states
$|S_z=-n_c/2\rangle$ and $|S_z=-n_c/2 +1\rangle$. Due to the 
$n_c+1$-fold degeneracy of the ground state
multiplet, the system is more
prone to leakage than the antiferromagnetic systems considered here
because quantum mechanical selection rules no longer prevent transitions
from the computational basis to other states in the ground state multiplet.

\subsection{XY-like chains}
\label{sec:chains3}

Our considerations have so far been focused on an isotropic exchange
coupling which is found, e.g., for coupled quantum dots.~\cite{burkard:99}
However, spin cluster qubits could  not only be realized in such
artificial superstructures. Rather, a wide variety of spin--$1/2$
chains in which magnetic ions are coupled by an exchange interaction 
has been synthesized.~\cite{mikeska:91} The control of single ion spins in
a molecule or solid in which the ions are rather closely spaced is even more
challenging than the control of a single spin in a quantum dot. Defining
a qubit in terms of the collective state of a short spin chain 
will make it substantially simpler to address qubits. However,
in contrast to coupled quantum dots, the exchange coupling in 
many spin chains found in nature is anisotropic.~\cite{mikeska:91}
Hence, we next turn to a discussion of anisotropic chains, $J_\perp \neq J_z$. 

For odd $n_c$, the spectrum still exhibits a ground state doublet of 
$\hat{S}_z$ eigenstates with eigenvalues $\pm \hbar/2$, respectively. 
However, $\hat{\bf S}^2$ is no longer a good quantum number.
Both for $J_\perp \gg J_z$ (XY-like systems) and $J_\perp \ll J_z$
(Ising-like systems), $|0\rangle$ and $|1\rangle$ can be explicitly 
constructed.

We first consider the XY model with $J_z=0$ in Eq.~(\ref{eq:sch-1}). By the
Jordan-Wigner transformation,~\cite{auerbach:94}
the XY chain is mapped onto a system of noninteracting spinless 
fermions on a lattice with spatially varying hopping amplitudes,
\begin{equation}
\hat{H} = - \frac{J_\perp}{2} \sum_{j=1}^{n_c-1} f_j 
(\hat{\psi}_{j+1}^\dagger \hat{\psi}_j +\hat{\psi}_{j}^\dagger 
\hat{\psi}_{j+1} ),
\label{eq:1p}
\end{equation}
where 
\begin{equation}
\hat{\psi}_j  = \exp \left[ i \pi  \sum_{k=1}^{j-1} \left( \hat{s}_{k,z} 
+ \frac{1}{2} \right) \right] \hat{s}_{j}^- \label{eq:jw1}
\end{equation}
annihilates a Jordan-Wigner fermion at site $j$.
The problem is thus reduced to calculating the one-particle eigenenergies
and eigenstates of Eq.~(\ref{eq:1p}). The one-particle Hamiltonian 
has $(n_c-1)/2$ pairs of states with negative and positive energy $\mp E_i$, 
respectively, which are related to each other by staggering of the wave 
function,
\begin{equation}
{\bf e}_i^{\pm} = \left(e_{i,1}, \pm e_{i,2}, e_{i,3}, \pm e_{i,5}, 
\ldots  \right),
\label{eq:xy-es}
\end{equation}
where $e_{i,j}$ are real numbers. In addition, there is one eigenstate 
\begin{equation}
{\bf e}_0 \propto \left(1,0,-\frac{f_1}{f_2},0,
\frac{f_1 f_3}{f_2 f_4},0, \ldots, \pm  
\frac{f_1 f_3 \ldots f_{n_c-2}}{f_2 f_4 \ldots f_{n_c-1}}\right)
\label{eq:xy-eigenstate}
\end{equation}
with energy eigenvalue $0$. The ground state doublet of the XY chain 
corresponds to the lowest $(n_c-1)/2$ 
and $(n_c+1)/2$ Jordan-Wigner fermion levels filled. Similarly to the
spin chain with isotropic exchange interactions, 
one-qubit gates can be realized by magnetic fields $B_z(t)$ and $B_x(t)$.
By numerical exact diagonalization of small spin chains ($n_c=9$),
we have shown that $|\langle 1|\hat{S}_x|0\rangle|$ remains of order $1/2$ 
for various set of $f_j$ [e.g., $f_j \equiv 1$, $f_j = \sin (j \pi/n_c)$], 
such that the operation time for the one-qubit rotation gate is 
only limited by  $\hbar/\Delta$. An isotropic interqubit coupling 
$\hat{H}_{\ast}=J_{\ast}(t) \hat{\bf s}_{n_c}^{\rm I} \cdot 
\hat{\bf s}_{1}^{\rm II}$ for
two-qubit gates still translates into the effective Hamiltonian in 
Eq.~(\ref{eq:cnot1}). 
With Eqs.~(\ref{eq:xy-es}) and (\ref{eq:xy-eigenstate}), from the 
completeness relation $\sum_{i=1}^{(n_c-1)/2} (
{e}_{i,j}^{+2} + {e}_{i,j}^{- 2}) + {e}_{0,j}^2 = 1$ 
for $j=1, \ldots, n_c$ and $\hat{s}_{j,z} = \hat{\psi}^\dagger_j \psi_j 
- 1/2$, one can calculate all matrix elements 
entering the effective coupling Hamiltonian Eq.~(\ref{eq:cnot1}),
\begin{eqnarray}
\langle 0| \hat{s}_{n_c,z}|0\rangle &=& \frac{{e}_{0,1}^2}{2}, 
\label{eq:xy-matrixel1} \\
|\langle 1 | \hat{s}_{n_c,x}|0\rangle | &=&  \frac{{e}_{0,1}}{2},
\label{eq:xy-matrixel2}
\end{eqnarray}
where ${e}_{0,1}$ is the first component of the normed 
one-particle eigenstate
defined in Eq.~(\ref{eq:xy-eigenstate}). In particular, for $f_j \equiv 1$, 
$\langle 0| \hat{s}_{n_c,z}|0\rangle = 1/(n_c+1)$ and 
$|\langle 1| \hat{s}_{n_c,x}|0\rangle| = 1/\sqrt{2(n_c+1)}$.
Due to the anisotropy of the intrachain exchange coupling, $\hat{H}_{\ast}$
(which is isotropic in the single-spin operators) translates into an 
effective XXZ-Hamiltonian in the two-qubit product basis. Nevertheless, 
the CNOT gate can still be realized according to 
Eq.~(\ref{eq:cnot2}). For the anisotropic chain, a magnetic field applied along
an axis ${\bf n}$ translates into a rotation around the axis 
$(n_x, n_y,n_z/2|\langle 1|\hat{S}_x|0\rangle|)$ in the Hilbert space 
spanned by $\{|0\rangle,|1\rangle\}$,
\begin{eqnarray}
\hat{H}^\prime &=& g \mu_B B {\bf n}\cdot \hat{\bf S} \nonumber \\
&\simeq & g \mu_B B 2 | \langle 1|\hat{S}_x |0 \rangle | \bigl[
n_x (|0\rangle\langle 1| + |1\rangle\langle 0|)/2 \nonumber \\
& & \hspace*{1cm} + 
n_y i (|1\rangle\langle 0| - |0\rangle\langle 1| )/2 \bigr]
\label{eq:xyeffham} \\
&& \hspace*{1cm} + g \mu_B B n_z(|0\rangle\langle 0| - |1\rangle\langle 1|)/2.
\nonumber
\end{eqnarray}
A one-qubit rotation around an
arbitrary axis, e.g. ${\bf n}_1$ in Eq.~(\ref{eq:cnot2}),
hence requires appropriate rescaling of the magnetic fields.~\cite{meier:02}

\subsection{Ising-like chains}
\label{sec:chains4}

\begin{figure}
\centerline{\mbox{\includegraphics[width=8cm]{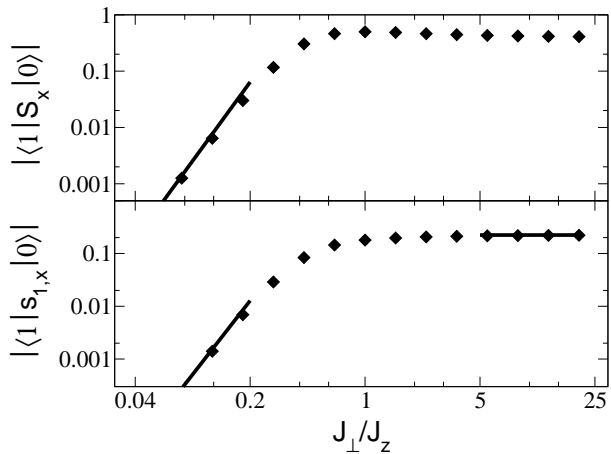}}}
\caption{Transition matrix elements $|\langle 1|\hat{S}_x|0\rangle|$ and
$|\langle 1|\hat{s}_{1,x}|0\rangle|$ as a function of exchange anisotropy
($J_\perp \neq J_z$). The matrix elements 
determine gate operation times for the one-qubit rotation and CNOT gate. 
Diamonds show numerical results obtained for  
$n_c = 9$ and spatially constant exchange interactions, $f_j \equiv 1$, 
in comparison with analytical results (solid lines) [see
Eq.~(\ref{eq:ising2})]. }
\label{Fig5}
\end{figure}

In the Ising limit $J_z \gg J_\perp$ the 
ground state doublet 
\begin{eqnarray}
|0\rangle &=&|\uparrow\rangle_1|\downarrow\rangle_2\ldots
|\uparrow\rangle_{n_c} + {\mathcal O}(J_\perp/J_z), \nonumber \\ 
|1\rangle &=& |\downarrow\rangle_1|\uparrow\rangle_2\ldots
|\downarrow\rangle_{n_c} + {\mathcal O}(J_\perp/J_z)
\label{eq:ising}
\end{eqnarray}
is~separated~from the next excited state by an $n_c$-independent 
$\Delta \sim J_z \min(f_j)$. In perturbation theory 
in $J_\perp/J_z$, for $f_j \equiv 1$, the matrix elements
\begin{eqnarray}
|\langle 1|\hat{S}_x|0\rangle| & \simeq &  
\frac{n_c+1}{4} \left( \frac{2J_\perp}{J_z} \right)^{(n_c-1)/2}, \nonumber \\
|\langle 1|\hat{s}_{n_c,x}|0\rangle| & \simeq & \frac{1}{2}
\left( \frac{2J_\perp}{J_z} \right)^{(n_c-1)/2}
\label{eq:ising2}
\end{eqnarray} 
decrease exponentially with system size because
$\hat{S}_x$ and $\hat{s}_{n_c,x}$ only flip one spin within 
the chain. Expanding the states $|0\rangle$ and $|1\rangle$ in powers of
$J_\perp/J_z$ it
follows that finite matrix element of $\hat{s}_{1,x}$ and $\hat{S}_x$ 
between $|0\rangle$ and $|1\rangle$ occur only in order 
$(n_c-1)/2$ in $J_\perp/J_z$.~\cite{villain:75}  
Even for medium sized chains $n_c \gtrsim 9$ 
and $J_\perp/J_z < 0.2$, an isotropic inter-qubit
coupling Hamiltonian $\hat{H}_{\ast}$ translates into an effective 
Hamiltonian Eq.~(\ref{eq:cnot1}) of Ising form (Fig.~\ref{Fig5}). 
Because of the long gate operation times implied by Eq.~(\ref{eq:ising2}) 
for the one-qubit rotation and, in particular, the CNOT gate, 
only quantum computing schemes which require a small number of such
operations~\cite{raussendorf:01} appear feasible. 

In Fig.~\ref{Fig5}, for a chain with $n_c=9$, we compare our analytical
results (solid lines) for the matrix elements 
$|\langle 1|\hat{S}_x|0\rangle|$ and 
$|\langle 1|\hat{s}_{1,x}|0\rangle|$ as functions of anisotropy
with exact diagonalization for the chain with $f_j \equiv 1$ (symbols). 
Because matrix elements of order unity imply quantum gate operation
times comparable to single spins, the results in Fig.~\ref{Fig5} show that 
universal quantum computing  based on a sequence of one-qubit rotation and 
CNOT gates is feasible for a wide range of  spin cluster qubits.

\subsection{Experimental realization}
\label{sec:chains6}

We illustrate the advantages of a spin cluster qubit formed by $n_c=5$ spins
$s=1/2$ in a one-dimensional array of quantum dots with diameter
$d \simeq 50$~nm.
For realistic parameters~\cite{burkard:99} 
($J \simeq 10$~K$k_B$, $J_\ast \simeq 2.3$~K$k_B$,
$g\mu_B B \simeq 0.7$~K$k_B$) 
and a magnetic field which decreases from its
maximum values $B_x$ at the central site of the chain to
$0.2 B_x$ at sites $j=2,4$, the operation
time for one-qubit gates increases by a factor $1/2|\langle 1|
 0.2 \hat{s}_{2,x}+ \hat{s}_{3,x} + 0.2 \hat{s}_{4,x} |0\rangle| \simeq 2.2$ 
compared to a  single spin. 
The operation time for the square-root of SWAP gate is increased by 
a factor of $1/(2|\langle 0|\hat{s}_{1,z}|0\rangle|)^2\simeq 4$ compared to
the single-spin qubit. However, the operation time for the CNOT gate as 
defined in Eq.~(\ref{eq:cnot2}) is mainly
determined by the single-qubit operations of the sequence. 
Hence, for the minimum operation time of the CNOT gate we find 
$386$~ps for spin clusters instead of $165$~ps for single spins. 
The decrease of decoherence time 
strongly depends on the microscopic origin of
decoherence. From the heuristic argument in Sec.~\ref{sec:chains1} 
we find that the decoherence rate 
due to globally fluctuating magnetic
fields does not scale with $n_c$ and is 
equal to the one of the single-spin qubit. Decoherence caused by fluctuating
local fields scales
linearly with $n_c$. For a spin cluster qubit with $n_c=5$, 
we estimate that the number of
quantum gates which can be performed during the decoherence time 
decreases by at most a factor of $10$ compared to the single-spin qubit. 
However, in contrast to single-spin qubits, magnetic fields and exchange 
constants 
must be controlled only over length scales $n_c d = 250$~nm and $2n_c d
= 500$~nm, respectively instead of $d=50$~nm. This would allow 
one to control the exchange interaction between  neighboring clusters 
optically.~\cite{levy:01} Note that, for the small clusters analyzed here, 
$\Delta$ is so large that neither adiabaticity nor the requirement that 
the Zeeman energy and $J_\ast$ be small compared to $\Delta$ provides a 
serious restriction on quantum gate operation times. 

Decoherence due to dipolar interactions is unimportant for electron spins
in quantum dots with a typical distance $d=50-100$~nm, where $E_{\rm dip}$
[Eq.~(\ref{eq:dip2})] is of order $10^{-11} {\rm K}k_B = \hbar/0.25 s$. 
However, for P dopants
in a Si matrix with a typical distance $d=100$~\AA, 
$E_{\rm dip} = 6.2 \times 10^{-7} {\rm K}k_B = \hbar/12\mu s$. Recent
experiments~\cite{tyryshkin:03} show that magnetic dipolar interactions
are indeed among the most dominant decoherence mechanisms for P dopants in a Si
matrix. The characteristic 
time scale for dephasing by dipolar interactions, $12\mu s$, is larger than
realistic gate operation times by only a factor 
$10^2-10^3$.~\cite{koiller:02,friesen:02} However, for a spin cluster formed by
$n_c=9$ spins with a dominant nearest neighbor exchange interaction, the 
intercluster dipolar interaction energy is reduced compared
to the one of single spins by a factor of $8$, i.e., the characteristic
decoherence time for dipolar interactions increases to $100 \mu s$.

\section{Spin clusters in $d>2$}
\label{sec:2d}

So far, our considerations have been restricted to spin chains. 
The main ideas discussed above apply to 
a much larger class of antiferromagnetic systems with 
uncompensated sublattices. We illustrate next 
that quantum gates are feasible also if spins $s=1/2$ are arranged in a 
two- or three-dimensional cluster. For definiteness, we restrict our
attention to an isotropic exchange interaction $J>0$.

\subsection{Bipartite lattice}
\label{sec:2d1}

\begin{figure}
\centerline{\mbox{\includegraphics[width=8cm]{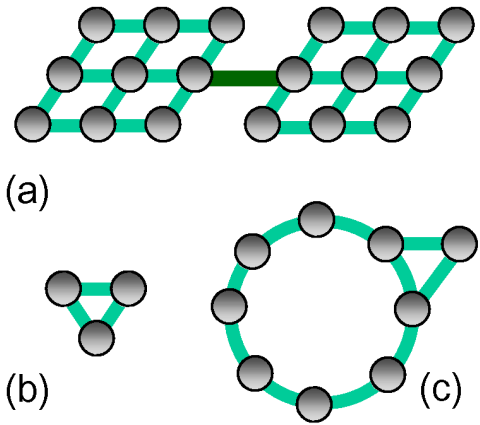}}}
\caption{Two-dimensional spin clusters. Each dot represents a single-spin 
qubit. (a) The spin cluster qubit scheme is readily extended from spin chains
to any bipartite lattice. (b) Spin arrays with frustrated bonds
have a highly degenerate ground state (fourfold degeneracy for three spins).
(c) If the frustrated bonds are part of a larger array, the high degeneracy is
usually lifted and a ground state doublet remains.}
\label{Fig6}
\end{figure}

We first consider an odd number of spins arranged
on a bipartite lattice with the number of sublattice sites 
differing by $1$, e.g., a rectangular lattice
with $L_1 \times L_2$ sites, where $L_{1,2}$ are odd. This two-dimensional 
lattice exhibits a ground state doublet~\cite{lieb:62}. Similarly to the 
spin chain, the computational basis $\{|0\rangle,|1\rangle\}$ can be 
defined in terms of the $\hat{S}_z$-eigenstates
of the ground state doublet. Here, $\hat{\bf S}$ is the operator of total spin
of the two-dimensional array. From a spin-wave ansatz for the elementary
excitations, the energy gap $\Delta$ separating the ground state doublet
from the next excited state can be estimated as 
\begin{equation}
\Delta \simeq \frac{J \pi}{\min(L_1,L_2)}. 
\label{eq:2dgap}
\end{equation}
Because the characteristic features of the ground state
doublet carry over from the one- to the two-dimensional spin cluster qubit,
quantum computing with two-dimensional spin arrays on bipartite lattices
is possible with the techniques discussed in Sec.~\ref{sec:chains1}. 
Gate operation times are determined by the matrix elements
$|\langle 0| \hat{s}_{j,z}|0\rangle|=|\langle 1| \hat{s}_{j,x}|0\rangle|$.
For the $3\times3$-lattice shown in Fig.~\ref{Fig6}(a), from exact 
diagonalization we find 
$\langle 0| \hat{s}_{j,z}|0\rangle = 0.15$ for sites $j$ in the center
of the edges, 
$\langle 0| \hat{s}_{j,z}|0\rangle = 0.23$ for sites $j$ at the corners,
and $\langle 0| \hat{s}_{j,z}|0\rangle = 0.17$ for the central site of
the cluster. Similarly to the spin chain, the ground state doublet is robust 
against a spatial variation in the exchange constants as
long as all exchange constants remain antiferromagnetic.

\subsection{Geometrically frustrated systems}
\label{sec:2d2}

For non-bipartite lattices, a ground-state doublet is not guaranteed to
emerge. The simplest example is the geometrically frustrated
system of three spins $s=1/2$ shown in Fig.~\ref{Fig6}(b),
\begin{eqnarray}
\hat{H} &=& J(\hat{\bf s}_1 \cdot\hat{\bf s}_2 +
\hat{\bf s}_1 \cdot\hat{\bf s}_3 + \hat{\bf s}_2 \cdot\hat{\bf s}_3) \nonumber
\\ &=& \frac{J}{2} \left(\hat{\bf S}^2 - \frac{9}{4} \right), 
\label{eq:frust1}
\end{eqnarray}
which has a fourfold degenerate ground state with 
energy eigenvalue $E=-0.75 J$. The eigenstates can be chosen as
\begin{eqnarray}
|0\rangle &=& \left(|\uparrow \uparrow \downarrow \rangle
-  |\uparrow  \downarrow  \uparrow\rangle \right)/\sqrt{2}, \nonumber \\
|1\rangle &=& \left(|\downarrow \uparrow \downarrow \rangle
-  |\downarrow  \downarrow  \uparrow\rangle \right)/\sqrt{2}, \nonumber \\
|2\rangle &=& \left(2 |\uparrow \downarrow \downarrow \rangle
-  |\downarrow  \uparrow  \downarrow\rangle 
- |\downarrow  \downarrow  \uparrow\rangle  \right)/\sqrt{6}, \label{eq:frust2}
\\
|3\rangle &=& \left(2 |\downarrow \uparrow \uparrow \rangle
-  |\uparrow  \uparrow  \downarrow\rangle 
- |\uparrow  \downarrow  \uparrow\rangle  \right)/\sqrt{6} .  \nonumber 
\end{eqnarray}
As demonstrated in Ref.~\onlinecite{kempe:01}, these states could still define 
a logical qubit robust against certain sources of decoherence. However,
quantum gate operation would always require control over
single exchange interactions or local magnetic fields and  
exclude quantum computing with control parameters which vary slowly in space.

Geometrical frustration does, however, not in general rule out 
the existence of a ground state doublet. In the more generic case that
geometrically frustrated bonds are part of a larger system 
[Fig.~\ref{Fig6}(c)]
or the exchange constants in Fig.~\ref{Fig6}(b) are not all equal, 
a ground state doublet emerges. In this case, the logical states of the spin 
cluster qubit
again can be defined in terms of the $\hat{S}_z$ eigenstates of the
ground state doublet and quantum gate operation 
is possible with magnetic fields and exchange constants varying slowly 
over the cluster. For systems as shown in Fig.~\ref{Fig6}(c), in which 
some bonds are frustrated, $\Delta$
is usually smaller than in the case of bipartite lattices. For example, for
 Fig.~\ref{Fig6}(c), $\Delta = 0.157 J$ compared to $\Delta = 0.991 J$
for Fig.~\ref{Fig6}(a). According to the adiabaticity requirement, the 
small gap limits gate operation times more severely for the system in 
Fig.~\ref{Fig6}(c).

\subsection{Experimental realization}
\label{sec:2d3}

Because spin cluster qubits emerge also in  two-dimensional regular spin 
arrays, spin cluster qubits can be arranged 
in a plane if the positions of single spins can be controlled as, e.g., 
for lateral quantum dots [Fig.~\ref{Fig6}(a)]. 
For a spin cluster qubit formed by $L\times L=n_c$ quantum
dots, $\Delta \propto 1/L$. Decoherence due to globally fluctuating fields
does not increase with $n_c$, whereas independent local Gaussian white noise
gives rise to a decoherence rate $1/\tau_\phi \propto n_c$. Two-dimensional
spin cluster qubits are hence particularly interesting for qubits
in which decoherence is induced mainly by global rather than local
fluctuating fields.

More importantly, a spin cluster qubit can be defined even for a wide range 
of systems
in which the positions and exchange constants cannot be accurately controlled.
For P atom electron spins in a Si matrix, because of rapid oscillations
of the exchange coupling between atoms at large distances, placement of 
atoms with lattice spacing precision is required for single-spin 
qubits.~\cite{koiller:02}  Without this precision, the exchange interaction 
at large distances vanishes with a probability of $50$\%.
In contrast, for spin cluster qubits formed by a small number (e.g. three) of 
P dopants located close to each other, the spin defining the logical state 
of the qubit is delocalized over the cluster. 
The effective exchange coupling between neighboring qubits obtained by 
integration of the exchange interaction over the clusters is 
finite with a high probability. Because the
intracluster exchange interaction at small distances varies strongly 
with distance,~\cite{koiller:02} for a random arrangement of three spins, 
the exchange couplings will differ with high
probability and the system is not frustrated.

\section{Larger spins}
\label{sec:ls}

So far, our considerations have been restricted to clusters formed
by spins $s=1/2$.
We next consider antiferromagnetic systems with larger spins $s>1/2$.

\subsection{Antiferromagnetic molecular clusters}
\label{sec:ls1}

Only very
recently it has been shown theoretically 
that Grover's algorithm can be implemented with
ferromagnetic molecular magnets using a unary 
encoding.~\cite{leuenberger:01,leuenberger:02}
In view of universal quantum computing, ferromagnetic clusters such 
as the molecular magnets  Mn$_{12}$ and Fe$_8$ 
(Ref.~\onlinecite{gatteschi:94}) suffer 
from the
large net spin which usually means large matrix elements coupling
the spin to the environment and, hence, short
decoherence times. 

In contrast, in antiferromagnetic systems, such as various ring
compounds,~\cite{gatteschi:94}
the spins couple such that they form 
a small total magnetic moment. Antiferromagnetic
clusters which have unequal sublattice magnetization,  
will in general have a ground 
state multiplet rather than the singlet found for systems with compensated
sublattice spins.~\cite{meier:01}
Several antiferromagnetic molecular magnets comprised of spins with 
quantum numbers larger than $1/2$ have been synthesized to 
date~\cite{taft:94,caneschi:96,waldmann:99,waldmann:01,slageren:02}
including several compounds with uncompensated sublattice 
spins.~\cite{larsen:03}
As a paradigm,
we consider systems with isotropic exchange interaction $J$, but allow for an
easy or hard axis single-spin  anisotropy,
\begin{equation}
\hat{H} = J \hat{\bf s}_1 \cdot \hat{\bf s}_2 + k_z (\hat{s}_{1,z}^2 + 
\hat{s}_{2,z}^2).
\label{eq:molmag}
\end{equation}
Here, $s_1$ and $s_2 = s_1 \pm 1/2$ are the spin quantum numbers of the
two sublattices, respectively, $J>0$ is an effective exchange constant, 
and $k_z$ the single ion anisotropy. Equation~(\ref{eq:molmag}) has a ground 
state doublet 
$\{|0\rangle,|1\rangle \}$ of $\hat{S}_z = \hat{s}_{1,z}+\hat{s}_{2,z}$ 
eigenstates with eigenvalues $\pm \hbar/2$, respectively.
Because $[\hat{S}_z,\hat{H}]=0$ 
for the Hamiltonian Eq.~(\ref{eq:molmag}), the logical qubit basis states
have an expansion of the form (for $s_2 = s_1-1/2$) 
\begin{eqnarray}
|0\rangle &=& \sum_{m_1=-s_1+1}^{s_1} \alpha_{m_1} |m_1,1/2- m_1\rangle 
\nonumber \\
&=& \alpha_{-s_1+1} |-s_1+1,s_1-1/2\rangle \label{eq:mm-rep}
 \\ && +  \alpha_{-s_1+2}
 |-s_1+2,s_1-3/2\rangle + \ldots \nonumber \\
&&   + \alpha_{s_1} |s_1,-s_1+1/2\rangle
\nonumber
\end{eqnarray}
in terms of the spin product basis.   
For $s_{1,2} \gg 1$, analytical expressions can be derived both for the 
action of 
a magnetic field (one-qubit rotation gate) and for the action of an 
interqubit coupling Hamiltonian (two-qubit gates) between clusters
I and II,
\begin{equation}
\hat{H}_\ast = J_\ast (t) \hat{\bf s}_2^{\rm I} \cdot \hat{\bf s}_1^{\rm II},
\label{eq:mm-cnot}
\end{equation}
within a coherent state path integral 
formalism.~\cite{klauder:79,auerbach:94} We only state the main
results of our calculations here. Further details  are given 
in Appendix~\ref{sec:ap3}.

\subsection{Hard axis systems}
\label{sec:ls2}

For strong hard axis anisotropy, $k_z>0$ and  $k_z s_{1,2}^2/ J \gg 1$,  
the spins ${\bf s}_1$ and ${\bf s}_2$ lie close to the $x$-$y$ plane for
both states of the ground state doublet.  
A large contribution in the expansion Eq.~(\ref{eq:mm-rep}) comes from the
states $|m_1 = 0\rangle |m_2 = 1/2 \rangle$ and
$|m_1 = 0\rangle |m_2 = -1/2 \rangle$, respectively. 
For illustration, for $s_1=3$, $s_2=5/2$, and
$k_z/J = 0.2$, by numerical diagonalization of Eq.~(\ref{eq:molmag}), we
find 
\begin{eqnarray}
|0\rangle &=& 0.25 \Bigl| 3,-\frac{5}{2} \Big\rangle  
- 0.41 \Bigl| 2,-\frac{3 }{2} \Big\rangle  \nonumber \\ && 
+ 0.52 \Bigl| 1,-\frac{1}{2} \Big\rangle 
- 0.52  \Bigl| 0,\frac{1}{2} \Big\rangle \label{eq:mm-rep2} \\ 
&& 
+0.42  \Bigl| -1,\frac{3}{2} \Big\rangle 
- 0.24  \Bigl| -2,\frac{5}{2} \Big\rangle. \nonumber
\end{eqnarray}
The state $|1\rangle$ is obtained by $|m_1,m_2\rangle \rightarrow
|-m_1,-m_2\rangle$ on the right hand side of Eq.~(\ref{eq:mm-rep2}).
In agreement with the semiclassical theory, a major contribution to
$\{|0\rangle,|1\rangle\}$ comes from  states with small $m_1$ and $m_2$.

In the following, we restrict our attention to systems with large anisotropy, 
$k_z (s_1^2+ s_2^2)/J \gg 1$. Then, $\Delta \simeq J$ (Appendix~\ref{sec:ap3})
and
\begin{eqnarray}
|\langle 1| \hat{S}_x |0\rangle| & = & 1/4, \nonumber \\
|\langle 1| \hat{s}_{1,x} |0\rangle| & = & s_1/2, 
\label{eq:mm-matrix} \\
|\langle 1| \hat{s}_{2,x} |0\rangle| & = & s_2/2. \nonumber
\end{eqnarray}
In particular, Eq.~(\ref{eq:mm-cnot}) translates into the effective 
Hamiltonian
\begin{eqnarray}
\hat{H}_\ast &=& J_\ast |\langle 0|\hat{s}_{1,z}|0\rangle| 
|\langle 0|\hat{s}_{2,z}|0\rangle| \left(
\begin{array}{cccc}
1 & 0 & 0 & 0 \\
0 & -1 & 0 & 0 \\
0 & 0 & -1 & 0 \\
0 & 0 & 0 & 1
\end{array}
\right) \nonumber \\ && \hspace*{1cm} + 
\frac{J_\ast s_1 s_2}{2} \left(
\begin{array}{cccc}
0 & 0 & 0 & 0 \\
0 & 0 & 1 & 0 \\
0 & 1 & 0 & 0 \\
0 & 0 & 0 & 0
\end{array}
\right)
\label{eq:mm2}
\end{eqnarray}
in the two-qubit product basis. As discussed
in Sec.~\ref{sec:chains1}, the CNOT gate can be
realized with a unitary time evolution governed by this effective qubit
coupling of the XXZ-form.

Matrix elements of order unity in Eq.~(\ref{eq:mm-matrix}) show that, e.g., 
a magnetic field $B_x$ efficiently
rotates the state $|0\rangle$ into $|1\rangle$. This is not a priori evident
given the rather complicated representation of the ground state doublet
in the single-spin product basis [Eq.~(\ref{eq:mm-rep})]. The large 
matrix elements arise because, for both $|0\rangle$ and 
$|1\rangle$, the spins lie close
to the $x$-$y$ plane in the hard axis system.  

\subsection{Easy axis systems}
\label{sec:ls3}

For $k_z<0$, configurations with spins aligned along the z-axis are
energetically favorable. 
We restrict our attention to systems with large anisotropy, $4 |k_z| (s_1^2
+ s_2^2)/J \gg 1$. Because a transition from $|0\rangle$ to $|1\rangle$ 
requires a rotation of both spins through a large energy barrier,
from the theory of spin quantum tunneling in antiferromagnetic 
systems~\cite{barbara:90,krive:90} we find that
$|\langle 1 | \hat{S}_x| 0 \rangle|, |\langle 1 | \hat{s}_{1,x}| 0 \rangle|,
|\langle 1 | \hat{s}_{2,x}| 0 \rangle| \propto
\exp(-\sqrt{8 |k_z| (s_1^2 + s_2^2)/J}) \ll 1$ are exponentially small. 
Similarly to a spin chain in the Ising limit [Sec.~\ref{sec:chains4}],
the easy axis system is a candidate for quantum
computing schemes as suggested in Ref.~\onlinecite{raussendorf:01}.

\begin{figure}
\centerline{\mbox{\includegraphics[width=8cm]{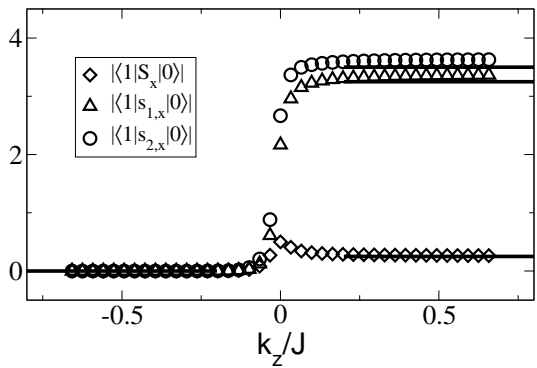}}}
\caption{Matrix elements of spin operators of a qubit formed by two
spins with spin quantum numbers $s_1=7$ and $s_2
= 6.5$. Numerical data (symbols) obtained from exact diagonalization
are in good agreement with analytical results (solid lines).
}
\label{Fig7}
\end{figure}

The analytical results for the matrix elements discussed here
are compared with numerical exact diagonalization for $s_1=7$ 
in Fig.~\ref{Fig7}. We find good agreement 
with our semiclassical results.

\subsection{Experimental realization}
\label{sec:ls4}

Single molecule
electrical switches~\cite{ball:00,collier:99,gittins:00,joachim:00} 
have nourished hopes that, in the
future, it will be possible to down-scale computers to the level at which 
bits or qubits are represented by single molecules. 
The results in Secs.~\ref{sec:ls2} and \ref{sec:ls3} show that, in such 
bottom-up approaches aiming at a universal quantum
computer, control is not required at the level of single atom
spins but only on the scale of molecule spins. In particular, molecular 
magnetic clusters with an effective spin $S=1/2$ define a qubit. 
One-qubit quantum gates could be effected, for example, by a magnetic tip as
used in magnetic force microscopy.~\cite{rugar:90} The spatial resolution
of these techniques currently lies in the range of 
$10-100$~nm~(Ref.~\onlinecite{liu:02}) and approaches the typical size of 
molecular magnetic clusters.~\cite{gatteschi:94}

Control of the exchange
interaction between molecules is challenging. 
As has been demonstrated recently,~\cite{collier:99,gittins:00}
the electrical conductivity of individual molecules can be switched
between two states in a controlled way. By connecting 
molecular magnetic clusters by reversible redox switches one could 
also switch intercluster exchange paths. Alternatively, if the relative
position of molecular magnetic clusters can be controlled, 
the intercluster exchange interaction can be switched on and off 
via the overlap of electron orbital wave functions by moving clusters
relative to each other.

\section{Conclusion}
\label{sec:conclusions}

In conclusion, we have shown that quantum computing is possible 
with a wide variety of clusters
assembled from antiferromagnetically coupled spins which form an 
effective total spin $S=1/2$. For arrays of spins
$s=1/2$, the existence of a spin cluster qubit requires 
little control over the placement and intracluster coupling of the 
spins and the spatial dimension of the array. This remains true 
for a wide range of systems with uncompensated sublattice spins 
differing by $1/2$. We have shown explicitly that, for the spin cluster
qubit, initialization, quantum gate operation, and readout are possible
with the techniques proposed and analyzed for single spins. The scaling
of the decoherence time with system size strongly depends on the 
microscopic decoherence mechanism. Spin cluster qubits are particularly 
promising in situations where decoherence is induced mainly by globally 
fluctuating fields during quantum gate operation and the decoherence rate
of the spin cluster qubit is comparable to that of a single-spin qubit or
for systems in which magnetic dipolar interactions are the dominant
decoherence mechanism. 
The main advantage of spin clusters compared to single
spins is that requirements on local control of magnetic fields and
exchange interactions can be traded for longer gate operation times.  
We have illustrated the feasibility and advantages of spin cluster qubits
for arrays of quantum dots, P dopants in a Si matrix, and molecular
magnetic clusters. 

In contrast to single spins, spin clusters are not intrinsically two-state
systems and leakage during quantum gate operation must
be accounted for. For the one-qubit rotation gate, we have shown
that leakage is small if the magnetic field which induces the rotation is
switched on and off adiabatically or if the Zeeman energy remains small
compared to $\Delta$. 

Finally, we note that, because any qubit can be mapped onto a spin $s=1/2$,
the results of this work do not only apply to quantum computing proposals
based on spin degrees of freedom but to any quantum computing scheme. 
More specifically, for any qubit for which methods for initialization,
quantum gate operation, quantum error correction, and readout have been 
identified, a cluster qubit
can be formed by coupling several qubits. For the
cluster qubit, initialization, quantum gate operation, quantum error 
correction, and readout are 
possible using the same techniques as for the original qubit.

\begin{acknowledgements}
This work was financially 
supported by the EU TMR network MOLNANOMAG, no. HPRN-CT-1999-00012 
(FM and DL), DARPA SPINS and QuIST 
(JL and DL), the Swiss NCCR Nanoscience (DL), and the Swiss NSF (DL). We
acknowledge discussions with G.~Burkard, V.~Cerletti, W.~Coish, H.~Gassmann, 
F.~Marquardt, and P.~Recher.\\ 
\end{acknowledgements}

\appendix

\section{Effective exchange Hamiltonian}
\label{sec:ap1}

Here, we derive explicitly the effective coupling Hamiltonian 
in Eq.~(\ref{eq:cnot1}) from Eq.~(\ref{eq:coup-ham}). The first and
second term in Eq.~(\ref{eq:cnot1}) result from 
$J_{\ast}(t) \hat{\bf s}_{n_c}^{\rm I} \cdot \hat{\bf s}_{1}^{\rm II}$
in the microscopic coupling. Decomposing
$\hat{\bf s}_{n_c}^{\rm I} \cdot \hat{\bf s}_{1}^{\rm II} =
\hat{s}_{n_c,z}^{\rm I} \hat{s}_{1,z}^{\rm II} + 
(\hat{s}_{n_c}^{{\rm I}+}\hat{s}_{1}^{{\rm II}-} 
+ \hat{s}_{n_c}^{{\rm I}-}\hat{s}_{1}^{{\rm II}+} )/2$ in terms of
spin ladder operators, one can readily evaluate the matrix elements
in the two-qubit product basis. Because, by definition, $\hat{S}_z |0\rangle
= (\hbar/2) |0\rangle$ and $\hat{S}_z |1\rangle
= (-\hbar/2) |1\rangle$  and
$\hat{s}_{n_c,z}^{\rm I} \hat{s}_{1,z}^{\rm II}$ conserves the 
$z$-component of total spin in each cluster separately it follows that
\begin{equation}
_{\rm I}\langle 0| \, _{\rm II} \langle 1| 
\hat{s}_{n_c,z}^{\rm I} \hat{s}_{1,z}^{\rm II} |1\rangle_{\rm I} 
|0\rangle_{\rm II} = 0.
\label{eq:me-det1}
\end{equation}
Similarly, all other off-diagonal elements of 
$\hat{s}_{n_c,z}^{\rm I} \hat{s}_{1,z}^{\rm II}$ vanish. 
Because of
\begin{equation}
 _{\rm I}\langle 0| \hat{s}_{n_c}^{{\rm I}\pm} |0\rangle_{\rm I}
= \, \, 
_{\rm I}  \langle 1| \hat{s}_{n_c}^{{\rm I}\pm} |1\rangle_{\rm I} = 0,
\label{eq:me-det2} 
\end{equation}
the transverse exchange interaction $J_{\ast}(t) 
(\hat{s}_{n_c}^{{\rm I}+}\hat{s}_{1}^{{\rm II}-} 
+ \hat{s}_{n_c}^{{\rm I}-}\hat{s}_{1}^{{\rm II}+} )/2$ 
has finite matrix elements only between the states
$|0\rangle_{\rm I} |1\rangle_{\rm II}$ and $|1\rangle_{\rm I} 
|0\rangle_{\rm II}$. This completes the proof that the intercluster
exchange term $J_\ast(t)  \hat{\bf s}_{n_c}^{\rm I} \cdot 
\hat{\bf s}_{1}^{\rm II}$ leads to the first  and second term in 
Eq.~(\ref{eq:cnot1}). 

It remains to show that a possible change in intracluster 
exchange interaction constants during two-qubit gate operation, 
$J_{\ast}(t) \sum_{j=1}^{n_c-1} (v_j^{\rm I}  \hat{\bf s}_{j}^{\rm I} \cdot
 \hat{\bf s}_{j+1}^{\rm I}+ v_j^{\rm II}  \hat{\bf s}_{j}^{\rm II} \cdot
 \hat{\bf s}_{j+1}^{\rm II})$, only leads to a term proportional to
${\bf 1}$ in 
Eq.~(\ref{eq:cnot1}). This term conserves all components of the
total spin of clusters I and II,
\begin{equation}
[\hat{S}_\alpha^{\rm I},  \sum_{j=1}^{n_c-1} 
v_j^{\rm I}  \hat{\bf s}_{j}^{\rm I} \cdot  \hat{\bf s}_{j+1}^{\rm I}] = 0,
\label{eq:me-det3}
\end{equation}
for $\alpha = x,y,z$, and similarly for II. Hence, all off-diagonal
matrix elements such as $_{\rm I}\langle 1 |
v_j^{\rm I}  \hat{\bf s}_{j}^{\rm I} \cdot \hat{\bf s}_{j+1}^{\rm I} 
|0 \rangle_{\rm I}$ vanish. Finally, because $|1\rangle_{\rm I}
= \hat{S}^{{\rm I}-} |0\rangle_{\rm I} = 2 \hat{S}_x^{\rm I} 
|0\rangle_{\rm I}$, with Eq.~(\ref{eq:me-det3}),
\begin{eqnarray}
&& _{\rm I}\langle 1| \, _{\rm II} \langle 0| 
J_{\ast}(t) \sum_{j=1}^{n_c-1} v_j^{\rm I}  \hat{\bf s}_{j}^{\rm I} \cdot
 \hat{\bf s}_{j+1}^{\rm I} |1\rangle_{\rm I} |0\rangle_{\rm II} \nonumber
\\
&& \hspace*{0.5cm}=  \, \, _{\rm I}  \langle 0| \, _{\rm II} \langle 0| 
2\hat{S}_x^{\rm I} 
J_{\ast}(t) \sum_{j=1}^{n_c-1} v_j^{\rm I}  \hat{\bf s}_{j}^{\rm I} \cdot
 \hat{\bf s}_{j+1}^{\rm I} 2\hat{S}_x^{\rm I}  
|0\rangle_{\rm I} |0\rangle_{\rm II} \label{eq:me-det4} \\
&& \hspace*{0.5cm} = \, \, _{\rm I} \langle 0| \, _{\rm II} \langle 0| 
J_{\ast}(t) \sum_{j=1}^{n_c-1} v_j^{\rm I}  \hat{\bf s}_{j}^{\rm I} \cdot
 \hat{\bf s}_{j+1}^{\rm I} (2\hat{S}_x^{\rm I})^2  
|0\rangle_{\rm I} |0\rangle_{\rm II} \nonumber  \\
&& \hspace*{0.5cm} = \, \, _{\rm I}   \langle 0| \, _{\rm II} \langle 0| 
J_{\ast}(t) \sum_{j=1}^{n_c-1} v_j^{\rm I}  \hat{\bf s}_{j}^{\rm I} \cdot
 \hat{\bf s}_{j+1}^{\rm I} 
|0\rangle_{\rm I} |0\rangle_{\rm II}. \nonumber
\end{eqnarray}
In the second line of Eq.~(\ref{eq:me-det4}) we have invoked that
$\{|0\rangle_{\rm I},|1\rangle_{\rm I} \}$ belong to one spin-$1/2$
doublet, the third line then follows from Eq.~(\ref{eq:me-det3}).
With a similar argument it can be shown that
all diagonal matrix elements in the two-qubit product basis are
equal and $J_{\ast}(t) \sum_{j=1}^{n_c-1} 
(v_j^{\rm I}  \hat{\bf s}_{j}^{\rm I} \cdot
\hat{\bf s}_{j+1}^{\rm I}+ v_j^{\rm II}  \hat{\bf s}_{j}^{\rm II} \cdot
\hat{\bf s}_{j+1}^{\rm II})$ translates  into a term
$J_o (t) {\bf 1}$ in the effective coupling Hamiltonian 
Eq.~(\ref{eq:cnot1}).

Finally, we prove Eq.~(\ref{eq:cnot1b}) which implies that,
for isotropic intracluster exchange interactions, the effective
two-qubit Hamiltonian is also of Heisenberg form. For simplicity,
we omit the label I of the spin cluster qubit in the following. 
In order to formally calculate $\langle 0 | \hat{s}_{n_c,z} |0\rangle$ and
$|\langle 1 | \hat{s}_{n_c,x} |0\rangle|$, we define the spin operators
\begin{equation}
\hat{S}^\prime_\alpha = \hat{S} - \hat{s}_{n_c,\alpha} = \sum_{j=1}^{n_c-1} 
\hat{s}_{j,\alpha} \label{eq:a-0}
\end{equation}
of all but the outermost spin $j=n_c$ of the cluster. Generally, 
$|0\rangle$ can be expanded as
\begin{equation}
|0\rangle = a | \Psi\rangle |\uparrow\rangle + b |\Phi\rangle |\downarrow 
\rangle,
\label{eq:a-1}
\end{equation}
where $| \Psi\rangle$ and $|\Phi\rangle$ describe the normed 
states of the leftmost
$n_c-1$ spins in the array and $a$ and $b$ are real numbers. 
Because $\hat{S}_z |0\rangle= (\hbar/2)|0\rangle$,
 $| \Psi\rangle$ and $|\Phi\rangle$ are eigenstates of $\hat{S}^\prime_z$ with
eigenvalues $0$ and $\hbar$, respectively. $\hat{S}_x |0\rangle
= (\hbar/2)|1\rangle$ is an $\hat{S}_z$ eigenstate with eigenvalue 
$-(\hbar/2)$, such that 
$b |\Phi\rangle = - a \hat{S}_+^\prime | \Psi\rangle$, and
\begin{eqnarray}
|0\rangle &=& a (| \Psi\rangle |\uparrow\rangle 
- \hat{S}_+^\prime | \Psi\rangle |\downarrow\rangle) \nonumber, \\
|1\rangle &=& a (\hat{S}_-^\prime| \Psi\rangle |\uparrow\rangle 
- (1- \hat{S}_-^\prime \hat{S}_+^\prime) 
| \Psi\rangle |\downarrow\rangle) \nonumber, \\
a &=& \frac{1}{\sqrt{1+ \langle \Psi|\hat{S}_-^\prime \hat{S}_+^\prime|\Psi
\rangle }}. \label{eq:a-2}
\end{eqnarray}
From Eq.~(\ref{eq:a-2}) we calculate
\begin{eqnarray}
&& \langle 0 | \hat{s}_{n_c,z} |0 \rangle 
= - \langle 1 | \hat{s}_{n_c,z} |1 \rangle = \frac{1}{2}
\frac{1-\langle \Psi|\hat{S}_-^\prime \hat{S}_+^\prime|\Psi\rangle}{1+\langle 
\Psi|\hat{S}_-^\prime \hat{S}_+^\prime|\Psi\rangle}, \nonumber \\
&& \langle 1| \hat{s}_{n_c,x} |0 \rangle =  \frac{1}{2}
\frac{1-\langle \Psi|\hat{S}_-^\prime \hat{S}_+^\prime|\Psi\rangle}{1+\langle 
\Psi|\hat{S}_-^\prime \hat{S}_+^\prime|\Psi\rangle}, \label{eq:a-3}
\end{eqnarray}
which proves Eq.~(\ref{eq:cnot1b}).

\section{Isotropic chain with spatially varying exchange interaction}
\label{sec:ap2}

The local spin density in the energy eigenstates of  
Eq.~(\ref{eq:sch-1}) depends sensitively on
spatial variations of the exchange interaction [Fig.~\ref{Fig2}(d)]. 
Whereas for
$f_j \equiv 1$ (solid lines) the magnetization density in each of the 
sublattices increases toward the center of the 
chain,~\cite{eggert:92} the opposite
behavior is observed for an exchange interaction $f_j = \sin(j\pi/n_c)$
(dashed lines). In the limit
$f_1=f_{n_c-1} \ll \min_{j=2,\ldots,n_c-2}f_j$, the increase of local spin 
density toward
the ends of the chain can be understood quantitatively. 
The ground state doublet
of the spin cluster qubit can be constructed explicitly from
the ground state doublet $\{|0\rangle_{n_c-2},|1\rangle_{n_c-2}\}$ of the
chain with the outermost spins removed. For $J f_1$ much smaller 
than the energy gap $\Delta_{n_c-2}$ of the chain formed by the $n_c-2$ 
central spins, the coupling of the outermost spins
can be treated perturbatively. For the chain with centrosymmetric
exchange couplings, $f_j = f_{n_c-j}$, from the ansatz
\begin{eqnarray}
|0\rangle  &=& 
\alpha_1 |\uparrow\rangle|1\rangle_{n_c-2}|\uparrow\rangle + 
\alpha_2 |\uparrow\rangle|0\rangle_{n_c-2}|\downarrow\rangle  \nonumber \\
&& 
+\alpha_3 |\downarrow\rangle|0\rangle_{n_c-2}|\uparrow\rangle +
{\mathcal O}(J f_1/\Delta_{n_c-2}),
\label{eq:weak-state}
\end{eqnarray}
we find $(\alpha_1,\alpha_2,\alpha_3) = (2,-1,-1)/\sqrt{6}$ 
for the ground state of the chain with $n_c$ spins and, hence,
\begin{equation}
\lim_{f_1 = f_{n_c-1}  \rightarrow 0}|\langle 0|\hat{s}_{1,z}|0\rangle|
=1/3.
\label{eq:weaklimit}
\end{equation} 

\section{Large spins}
\label{sec:ap3}

The matrix elements in  Eq.~(\ref{eq:mm-matrix}) 
can be calculated from
coherent state spin path integrals.~\cite{klauder:79} We focus on 
strong easy plane systems, $k_z>0$ and
$k_z(s_1^2 + s_2^2)/J \gg 1$. Following the standard approach for
antiferromagnetic systems, the partition function of the
two-spin system is expressed 
as path integral over the N{\'e}el vector ${\bf n}$ and
homogeneous magnetization ${\bf l}$ defined by ${\bf s}_1 = s_1 {\bf n}
+ {\bf l}$ and ${\bf s}_2 = - s_2 {\bf n}+ {\bf l}$, where
${\bf n}\cdot {\bf l}=0$. Integrating
out ${\bf l}$ in a saddle point approximation and parametrizing 
\begin{equation}
{\bf n} =  \left( \begin{array}{c} \sin \theta \, \cos \phi \\
\sin \theta \, \sin \phi \\
\cos \theta
 \end{array}  \right), \label{eq:ap2-param}
\end{equation}
the Euclidean action of the system can be written as~\cite{barbara:90,krive:90}
\begin{eqnarray}
{ L}_E = \frac{\hbar^2}{2J}\left(\dot{\theta}^2 + \sin^2 \theta \,
\dot{\phi}^2 \right) + k_z (s_1^2 + s_2^2) \cos^2 \theta \nonumber \\
+ i \Delta s \, \hbar \, \dot{\phi}(1-\cos \theta),
\label{eq:ap2-act1}
\end{eqnarray}
where the last factor accounts for the difference $\Delta s = 
s_2-s_1$ of the spin quantum numbers and $\dot{\phi} = \partial_\tau \phi$
is the imaginary time derivative. In the limit of strong
anisotropy, $k_z(s_1^2 + s_2^2)/J \gg 1$, Eq.~(\ref{eq:ap2-act1}) can be
expanded to second order in $\theta-\pi/2$ and the fluctuations are
integrated out, leading to
\begin{equation}
Z = \int {\mathcal D}\phi \, \exp \left(-\int_0^{\hbar \beta}
d \tau \, { L}_E[\phi]/\hbar \right) \label{eq:ap2-partfun}
\end{equation}
with an effective Euclidean Lagrangean
\begin{equation}
{ L}_E[\phi] = \frac{\hbar^2 \dot{\phi}^2}{2J} + i \Delta s \, \hbar \,
\dot{\phi}.
\label{eq:ap2-lagrange}
\end{equation}
After continuation to real time, by a canonical transformation we obtain the
Hamiltonian of the system in terms of the N{\'e}el vector operator,
\begin{equation}
\hat{H} = \frac{J}{2 \hbar^2} \left(\hat{p}_\phi - \hbar \Delta s \right)^2,
\label{eq:ap2-hamilt}
\end{equation}
where 
\begin{equation}
\hat{\bf n} = \left( \begin{array}{c} \cos \hat{\phi} \\ \sin \hat{\phi} \\
0 \end{array} \right), \,\, \hat{\bf l} = \frac{1}{2}
\left( \begin{array}{c} 0 \\ 0 \\
\hat{p}_\phi/\hbar - \Delta s \end{array} \right), \label{eq:ap2-spins}
\end{equation}
and $\hat{p}_\phi$ is the momentum operator conjugate to the in-plane 
polar angle $\phi$, $[\hat{p}_\phi,\hat{\phi}] = - i\hbar$. 
By inspection of Eq.~(\ref{eq:ap2-hamilt}), we find that the spin system 
Eq.~(\ref{eq:molmag}) has been mapped onto the Hamiltonian of a particle
on a ring threaded by a magnetic flux $\propto \Delta s$. In particular,
for half-integer $\Delta s$, the Hamiltonian has a ground state doublet
$\{|0\rangle,|1\rangle \}$ with wave functions 
$\psi_0(\phi) = \exp(i (m+1) \phi)/\sqrt{2 \pi}$ and $\psi_1(\phi) =
\exp(i m \phi)/\sqrt{2 \pi}$, where $m = \lfloor \Delta s \rfloor$. From 
$\hat{s}_{1,x} \simeq s_1 \cos \hat{\phi}$ and $\hat{s}_{2,x} \simeq 
- s_2 \cos \hat{\phi}$, one
immediately obtains Eq.~(\ref{eq:mm-matrix}).

\end{document}